\documentclass[longauth]{aa} 
\usepackage[pdftex]{graphicx}         
\usepackage{epstopdf}

\usepackage[hyphens]{url}
\usepackage{color}

\usepackage{txfonts}
\usepackage[authoryear,sort]{natbib}

\begin{document}
   \title{The 2009 multiwavelength campaign on Mrk 421: Variability and correlation 
   studies\thanks{The complete data set shown in Fig. \ref{fig:lightcurves} is 
   available in electronic form at the CDS via anonymous ftp to cdsarc.u-strasbg.fr 
   (130.79.128.5) or via http://cdsweb.u-strasbg.fr/cgi-bin/qcat?J/A+A/}}

\author{
\small
J.~Aleksi\'c\inst{1} \and
S.~Ansoldi\inst{2} \and
L.~A.~Antonelli\inst{3} \and
P.~Antoranz\inst{4} \and
A.~Babic\inst{5} \and
P.~Bangale\inst{6} \and
U.~Barres de Almeida\inst{6} \and
J.~A.~Barrio\inst{7} \and
J.~Becerra Gonz\'alez\inst{8} \and
W.~Bednarek\inst{9}
% ...
\and
\\
(The MAGIC collaboration) 
\\
S.~Archambault\inst{33} \and
B.~Behera\inst{10} \and
M.~Beilicke\inst{35} \and
W.~Benbow\inst{36} \and
R.~Bird\inst{37} \and
J.~H.~Buckley\inst{35} \and
V.~Bugaev\inst{35} \and
M.~Cerruti\inst{36} \and
X.~Chen\inst{38, 34} \and
L.~Ciupik\inst{39} \and
E.~Collins-Hughes\inst{37} \and
W.~Cui\inst{40} \and
J.~Dumm\inst{41} \and
J.~D.~Eisch\inst{42} \and
A.~Falcone\inst{43} \and
S.~Federici\inst{34, 38} \and
Q.~Feng\inst{40} \and
J.~P.~Finley\inst{40} \and
H.~Fleischhack\inst{10} \and
P.~Fortin\inst{59} \and
L.~Fortson\inst{41} \and
A.~Furniss\inst{44} \and
S.~Griffin\inst{33} \and
S.~T.~Griffiths\inst{45} \and
J.~Grube\inst{39} \and
G.~Gyuk\inst{39} \and
D.~Hanna\inst{33} \and
J.~Holder\inst{46} \and
G.~Hughes\inst{10} \and
T.~B.~Humensky\inst{47} \and
C.~A.~Johnson\inst{44} \and
P.~Kaaret\inst{45} \and
M.~Kertzman\inst{48} \and
Y.~Khassen\inst{37} \and
D.~Kieda\inst{49} \and
H.~Krawczynski\inst{35} \and
F.~Krennrich\inst{42} \and
S.~Kumar\inst{46} \and
M.~J.~Lang\inst{50} \and
G.~Maier\inst{10} \and
S.~McArthur\inst{51} \and
K.~Meagher\inst{52} \and
P.~Moriarty\inst{53, 50} \and
R.~Mukherjee\inst{54} \and
R.~A.~Ong\inst{55} \and
A.~N.~Otte\inst{52} \and
N.~Park\inst{51} \and
A.~Pichel\inst{56} \and
M.~Pohl\inst{38, 34} \and
A.~Popkow\inst{55} \and
H.~Prokoph\inst{10} \and
J.~Quinn\inst{37} \and
K.~Ragan\inst{33} \and
J.~Rajotte\inst{33} \and
P.~T.~Reynolds\inst{57} \and
G.~T.~Richards\inst{52} \and
E.~Roache\inst{36} \and
A.~C.~Rovero\inst{56} \and
G.~H.~Sembroski\inst{40} \and
K.~Shahinyan\inst{41} \and
D.~Staszak\inst{33} \and
I.~Telezhinsky\inst{38, 34} \and
M.~Theiling\inst{40} \and
J.~V.~Tucci\inst{40} \and
J.~Tyler\inst{33} \and
A.~Varlotta\inst{40} \and
S.~P.~Wakely\inst{51} \and
T.~C.~Weekes\inst{36} \and
A.~Weinstein\inst{42} \and
R.~Welsing\inst{10} \and
A.~Wilhelm\inst{38, 34} \and
D.~A.~Williams\inst{44} \and
B.~Zitzer\inst{58} \and
\\
(The VERITAS collaboration)
\\
M.~Villata\inst{60} \and 
C.~Raiteri\inst{60} \and 
H.~D.~Aller\inst{61} \and 
M.~F.~Aller\inst{61} \and 
W.~P.~Chen\inst{62} \and 
B.~Jordan\inst{63} \and 
E.~Koptelova\inst{62,64} \and 
O.~M.~Kurtanidze\inst{65,66} \and 
A.~L\"ahteenm\"aki\inst{67,68} \and 
B.~McBreen\inst{69} \and 
V.~M.~Larionov\inst{70,71,72} \and 
C.~S.~Lin\inst{62} \and 
M.~G.~Nikolashvili\inst{65} \and 
E.~Angelakis\inst{73} \and 
M.~Capalbi\inst{86} \and 
A.~Carrami\~nana\inst{75} \and 
L.~Carrasco\inst{75} \and
P.~Cassaro\inst{76} \and
A.~Cesarini\inst{77} \and 
L.~Fuhrmann\inst{73} \and
M.~Giroletti\inst{91} \and
T.~Hovatta\inst{78} \and 
T.~P.~Krichbaum\inst{73} \and 
H.~A.~Krimm\inst{79,80} \and 
W. Max-Moerbeck\inst{78} \and 
J.~W.~Moody\inst{83} \and 
G.~Maccaferri\inst{84} \and 
Y.~Mori\inst{85} \and 
I.~Nestoras\inst{73} \and 
A.~Orlati\inst{84} \and 
C.~Pace\inst{81} \and 
R.~Pearson\inst{83} \and 
M.~Perri\inst{3,86} \and 
A.C.S.~Readhead\inst{78} \and 
J.L.~Richards\inst{87} \and 
A.~C.~Sadun\inst{88} \and 
T.~Sakamoto\inst{89} \and 
J.~Tammi\inst{67,68} \and 
M.~Tornikoski\inst{67} \and 
Y.~Yatsu\inst{85} \and
A.~Zook\inst{90}
          }
\institute { IFAE, Edifici Cn., Campus UAB, E-08193 Bellaterra, Spain
\and Universit\`a di Udine, and INFN Trieste, I-33100 Udine, Italy
\and INAF National Institute for Astrophysics, I-00136 Rome, Italy
\and Universit\`a  di Siena, and INFN Pisa, I-53100 Siena, Italy
\and Croatian MAGIC Consortium, Rudjer Boskovic Institute, University of Rijeka and University of
	Split, HR-10000 Zagreb, Croatia
\and Max-Planck-Institut f\"ur Physik, D-80805 M\"unchen, Germany
\and Universidad Complutense, E-28040 Madrid, Spain
\and Inst. de Astrof\'isica de Canarias, E-38200 La Laguna, Tenerife, Spain
\and University of Lodz, PL-90236 Lodz, Poland
\and Deutsches Elektronen-Synchrotron (DESY), D-15738 Zeuthen, Germany
\and ETH Zurich, CH-8093 Zurich, Switzerland
\and Universit\"at W\"urzburg, D-97074 W\"urzburg, Germany
\and Centro de Investigaciones Energ\'eticas, Medioambientales y Tecnol\'ogicas, E-28040 Madrid, 
	Spain
\and Technische Universit\"at Dortmund, D-44221 Dortmund, Germany
\and Inst. de Astrof\'isica de Andaluc\'ia (CSIC), E-18080 Granada, Spain
\and Universit\`a di Padova and INFN, I-35131 Padova, Italy
\and Universit\`a dell'Insubria, Como, I-22100 Como, Italy
\and Unitat de F\'isica de les Radiacions, Departament de F\'isica, and CERES-IEEC, Universitat 
	Aut\`onoma de Barcelona, E-08193 Bellaterra, Spain
\and Institut de Ci\`encies de l'Espai (IEEC-CSIC), E-08193 Bellaterra, Spain
\and Finnish MAGIC Consortium, Tuorla Observatory, University of Turku and Department of
	Physics, University of Oulu, Finland
\and Japanese MAGIC Consortium, Division of Physics and Astronomy, Kyoto University, Japan
\and Inst. for Nucl. Research and Nucl. Energy, BG-1784 Sofia, Bulgaria
\and Universitat de Barcelona (ICC/IEEC), E-08028 Barcelona, Spain
\and Universit\`a di Pisa, and INFN Pisa, I-56126 Pisa, Italy
\and now at Ecole polytechnique f\'ed\'erale de Lausanne (EPFL), Lausanne, Switzerland
\and now at Department of Physics \& Astronomy, UC Riverside, CA 92521, USA
\and now at Finnish Centre for Astronomy with ESO (FINCA), Turku, Finland
\and also at INAF-Trieste
\and also at Instituto de Fisica Teorica, UAM/CSIC, E-28049 Madrid, Spain
\and now at Stockholm University, Fysikum, Oskar Klein Centre, AlbaNova, SE-106 91 
	Stockholm, Sweden
\and now at GRAPPA Institute, University of Amsterdam, 1098XH Amsterdam, Netherlands
 \and now at Stockholm University, Department of Astronomy, Oskar Klein Centre, AlbaNova, 
 	SE-106 91 Stockholm, Sweden, 
\and Physics Department, McGill University, Montreal, QC H3A 2T8, Canada
\and DESY, Platanenallee 6, 15738 Zeuthen, Germany
\and Department of Physics, Washington University, St. Louis, MO 63130, USA
\and Fred Lawrence Whipple Observatory, Harvard-Smithsonian Center for Astrophysics, Amado, AZ 85645, USA
\clearpage
\and School of Physics, University College Dublin, Belfield, Dublin 4, Ireland
\and Institute of Physics and Astronomy, University of Potsdam, 14476 Potsdam-Golm, Germany
\and Astronomy Department, Adler Planetarium and Astronomy Museum, Chicago, IL 60605, USA
\and Department of Physics and Astronomy, Purdue University, West Lafayette, IN 47907, USA
\and School of Physics and Astronomy, University of Minnesota, Minneapolis, MN 55455, USA
\and Department of Physics and Astronomy, Iowa State University, Ames, IA 50011, USA
\and Department of Astronomy and Astrophysics, 525 Davey Lab, Pennsylvania State University, University Park, PA 16802, USA
\and Santa Cruz Institute for Particle Physics and Department of Physics, University of California, Santa Cruz, CA 95064, USA
\and Department of Physics and Astronomy, University of Iowa, Van Allen Hall, Iowa City, IA 52242, USA
\and Department of Physics and Astronomy and the Bartol Research Institute, University of Delaware, Newark, DE 19716, USA
\and Physics Department, Columbia University, New York, NY 10027, USA
\and Department of Physics and Astronomy, DePauw University, Greencastle, IN 46135-0037, USA
\and Department of Physics and Astronomy, University of Utah, Salt Lake City, UT 84112, USA
\and School of Physics, National University of Ireland Galway, University Road, Galway, Ireland
\and Enrico Fermi Institute, University of Chicago, Chicago, IL 60637, USA
\and School of Physics and Center for Relativistic Astrophysics, Georgia Institute of Technology, 837 State Street NW, Atlanta, GA 30332-0430
\and Department of Life and Physical Sciences, Galway-Mayo Institute of Technology, Dublin Road, Galway, Ireland
\and Department of Physics and Astronomy, Barnard College, Columbia University, NY 10027, USA
\and Department of Physics and Astronomy, University of California, Los Angeles, CA 90095, USA
\and Instituto de Astronomia y Fisica del Espacio, Casilla de Correo 67 - Sucursal 28, (C1428ZAA) Ciudad Autonoma de Buenos Aires, Argentina
\and Department of Applied Physics and Instrumentation, Cork Institute of Technology, Bishopstown, Cork, Ireland
\and Argonne National Laboratory, 9700 S. Cass Avenue, Argonne, IL 60439, USA
\and Harvard-Smithsonian Center for Astrophysics, Cambridge, MA 02138, USA
\and INAF, Osservatorio Astronomico di Torino, I-10025 Pino Torinese (TO), Italy
\and Department of Astronomy, University of Michigan, Ann Arbor, MI 48109-1042, USA
\and Graduate Institute of Astronomy, National Central University, Jhongli 32054, Taiwan
\and School of Cosmic Physics, Dublin Institute for Advanced Studies, Dublin, 2, Ireland
\and Moscow M.V. Lomonosov State University, Sternberg Astronomical Institute, Russia
\and Abastumani Observatory, Mt. Kanobili, 0301 Abastumani, Georgia
\and Landessternwarte, Zentrum f\"{u}r Astronomie der Universit\"{a}t
Heidelberg,  K\"{o}nigstuhl 12, 69117 Heidelberg, Germany 
\and Aalto University Mets\"ahovi Radio Observatory Mets\"ahovintie 114 FIN-02540 Kylm\"al\"a Finland
\and Aalto University Department of Radio Science and Engineering, P.O.Box 13000, 
FI-00076 Aalto, Finland
\and University College Dublin, Belfield, Dublin 4, Ireland
\and Isaac Newton Institute of Chile, St. Petersburg Branch,
St. Petersburg, Russia
\clearpage
\and Pulkovo Observatory, 196140 St. Petersburg, Russia
\and Astronomical Institute, St. Petersburg State University, St. Petersburg, Russia
\and Max-Planck-Institut f\"ur Radioastronomie, Auf dem H\"ugel 69, 53121 Bonn, Germany
\and Agenzia Spaziale Italiana (ASI) Science Data Center, I-00044 Frascati (Roma), Italy
\and Instituto Nacional de Astrof\'isica, \'Optica y Electr\'onica, Tonantzintla, Puebla 72840, Mexico
\and INAF Istituto di Radioastronomia, Sezione di Noto,Contrada Renna Bassa, 96017 Noto (SR), Italy
\and Department of Physics, University of Trento, I38050, Povo, Trento, Italy
\and Cahill Center for Astronomy and Astrophysics, California Institute of Technology, 1200~E~California Blvd, Pasadena, CA 91125
\and Astro Space Center of the Lebedev Physical Institute, 117997 Moscow, Russia
\and Center for Research and Exploration in Space Science and Technology (CRESST) and NASA Goddard Space Flight Center, Greenbelt, MD 20771, USA
\and Indiana University, Department of Astronomy, Swain Hall West 319, Bloomington, IN 47405-7105, USA 
\and National Radio Astronomy Observatory, PO Box 0, Socorro, NM 87801
\and Department of Physics and Astronomy, Brigham Young University, Provo, Utah 84602, USA
\and INAF Istituto di Radioastronomia, Stazione Radioastronomica di Medicina, I-40059 Medicina (Bologna), Italy
\and Department of Physics, Tokyo Institute of Technology, Meguro City, Tokyo 152-8551, Japan
\and ASI-Science Data Center, Via del Politecnico, I-00133 Rome, Italy
\and Department of Physics, Purdue University, 525 Northwestern Ave, West Lafayette, IN  47907
\and Department of Physics, University of Colorado, Denver, CO 80220, USA
\and Department of Physics and Mathematics, College of Science and 952 Engineering, Aoyama Gakuin University, 5-10-1 Fuchinobe, Chuoku, Sagamihara-shi Kanagawa 252-5258, Japan
\and Department of Physics and Astronomy, Pomona College, Claremont CA 91711-6312, USA
\and INAF Istituto di Radioastronomia, 40129 Bologna, Italy
\and {*} Corresponding authors: Nina Nowak (nina.nowak@astro.su.se) and David Paneque (dpaneque@mppmu.mpg.de),        
}

   \date{}

  \abstract
   {
}
   {We perform an extensive characterization of the broadband
     emission of Mrk\,421, as well as its temporal evolution, during
     the non-flaring (low) state. The high brightness and nearby location ($z$=0.031) of Mrk\,421 make
     it an excellent laboratory to study blazar
     emission. The goal is to learn about the physical
     processes responsible for the typical emission of Mrk\,421, which
     might also be extended to other blazars that are located farther
     away and hence are more difficult to study.}
{We performed a 4.5-month 
   multi-instrument campaign on Mrk\,421 between January 2009 and June
   2009, which included VLBA, F-GAMMA, GASP-WEBT, \emph{Swift}, RXTE,
   \emph{Fermi}-LAT, MAGIC, and Whipple, among other instruments and
   collaborations. This extensive radio to very-high-energy (VHE;
   $E>100$~GeV) $\gamma$-ray dataset
   provides excellent temporal and energy coverage, which allows
   detailed studies of the evolution of the broadband spectral
   energy distribution.}
{Mrk421 was found in its typical
   (non-flaring) activity state, with a  VHE flux of about half that of the Crab Nebula, yet the
   light curves show significant variability at all wavelengths, the
   highest variability being in the X-rays. We determined the power spectral densities
   (PSD) at most wavelengths and found that all PSDs can be described
   by power-laws without a break, and with indices 
   consistent with pink/red-noise behavior. We observed a
   harder-when-brighter behavior 
   in the X-ray spectra and measured a positive
   correlation between VHE and X-ray fluxes with zero time lag. Such characteristics
   have been reported many times during flaring activity,
   but here they are reported for the first time in the non-flaring 
   state. We also observed an overall anti-correlation between
   optical/UV and X-rays extending over the duration of the campaign.}
   {The harder-when-brighter behavior in the X-ray spectra and the measured positive 
   X-ray/VHE correlation during the 2009 multi-wavelength
campaign suggests that the physical processes dominating the
emission during non-flaring states have similarities with those
occurring during flaring activity. In particular, this observation
supports leptonic scenarios as being responsible for the emission of
Mrk\,421 during non-flaring activity. Such a temporally
extended X-ray/VHE correlation is not driven by any single flaring
event, and hence is difficult to explain within the standard hadronic
scenarios. The highest variability is observed in the X-ray band, which, within the one-zone
synchrotron self-Compton scenario, indicates that the electron energy distribution is most
variable at the highest energies.}



   \maketitle
%

\defcitealias{Abdo-2011a}{Paper~I}
\section{Introduction}
Blazars are a class of radio-loud active galactic nuclei (AGN) where the relativistic
jet is believed to be closely aligned to our line of sight. They emit
radiation over a broad energy range from radio to very high energy
$\gamma$ rays (VHE; $E>100$~GeV), which is highly variable at all
wavelengths. Their spectral energy distributions (SED) are dominated
by the jet emission and show two bumps, one at low energies (radio,
optical, X-rays) and the other at high energies (X-rays,
$\gamma$ rays, VHE). While the origin of the low-energy bump is
presumably synchrotron emission from relativistic electrons, the
origin of the high-energy bump is still under debate. To constrain
current theoretical models for broadband blazar emission, simultaneous
observations of those objects over the whole wavelength range and over
a long period are needed. It is important to perform observations at
typical\footnote{We use the term ``typical'' instead of ``quiescent'',
to describe a state that is neither exceptionally high/flaring, nor
at the lowest possible level. Even though the term ``quiescent'' has
been used in the past to denote non-flaring activity in Mrk\,421 and
other blazars, we note that the term quiescent refers to the lowest
possible emission, which is actually unknown, and hence not suitable
in this context.} or even lower states in order to have a baseline
to which other (flaring) states can be compared, as distinct
physical processes might play a role when the source is flaring.  Weak
blazars in a low state are particularly poorly studied in $\gamma$ rays
because of the difficulty to detect them at these energies with
current instrumentation. In addition, most multi-wavelength programs
are triggered when a source is flaring, and not when it is in low
state.

The high-energy peaked BL Lac object (HBL) Mrk\,421 was the first
extragalactic object discovered at VHE \citep{Punch-1992}. It
is one of the brightest extragalactic X-ray/VHE objects, and because of 
its proximity ($z=0.031$) the absorption by the extragalactic
background light (EBL) is low \citep{Albert-2007}. Mrk\,421 has been
well-studied during phases of high activity, but simultaneous
broadband observations in a low state, covering both energy bumps, were
missing until recently.

Starting in 2009, a multi-wavelength (from radio to
VHE), multi-instrument  was organized to
monitor the broadband emission of Mrk\,421. The scientific goal was to
collect a complete, unbiased and simultaneous multi-wavelength dataset
to test current theoretical models of broadband blazar emission. In
this paper we analyze the temporal variability of Mrk\,421 in all
wavelengths during the 4.5-month observation period in 2009. During
the entire period, Mrk\,421 did not show any major flaring activity
\citep[e.g.,][]{Gaidos-1996, 2011ApJ...733...14M, Fossati-2008, Aleksic-2012, Fortson-2012}. The
multi-wavelength dataset is used to enhance our understanding of the
origin of the high-energy emission of blazars beyond the usually
observed flaring states. The underlying physical mechanisms
responsible for the acceleration of particles in jets are compared
with those observed during flares. This paper can be understood as a
sequel to \citet{Abdo-2011a} \citepalias{Abdo-2011a}, where the
SED of the Mrk\,421 2009 data was
analyzed.

This paper is organized as follows: Sect. \ref{sec:observations}
introduces the participating instruments and the multi-wavelength
data. The analysis of the variability in each waveband is presented in
Sect. \ref{sec:variability}. Cross-correlations and periodic
behavior are examined in Sects. \ref{sec:correlations} and
\ref{sec:periodicities}, and finally in Sect. \ref{sec:results} we
summarize and discuss our results.

\defcitealias{Pichel-2009}{Pichel (2009)}

\section{The 2009 multi-wavelength campaign}\label{sec:observations}

The duration of the 2009 campaign on Mrk\,421 was 4.5 months from 2009
January 19 (MJD~54850) to 2009 June 1 (MJD~54983).  29 instruments
participated in the campaign. The intended sampling was one
observation per instrument every two days, whenever weather, 
technical and observational limitations allowed\footnote{e.g., for imaging air Cherenkov 
telescopes (IACTs) like MAGIC or Whipple, observations during moonlight are only 
possible to a very limited extent, resulting in regular gaps of $\sim10$~days in the VHE 
light curves.}. The list of participating instruments
and the time coverage as a function of energy range are shown in Table 2
and in Figure 6 of \citetalias{Abdo-2011a}. The schedule of the observations
can be found
online\footnote{\url{https://confluence.slac.stanford.edu/display/GLAMCOG/Campaign+on+Mrk421+(Jan+2009+to+May+2009)}}.
The individual datasets and the data reduction are presented in
detail in section 5 of \citetalias{Abdo-2011a} and will therefore not
be introduced again in this paper. Besides the datasets reported in
Paper~I, this paper also reports VHE data from 115~hours of
dedicated Mrk\,421 observations with the Whipple 10-meter telescope (operated by
the VERITAS collaboration). These data are essential for the excellent temporal coverage 
in the VHE for this campaign. 
Details on the light curve presented here can 
be found in \citetalias{Pichel-2009} with
the general Whipple analysis technique described in \citet{Horan-2007} and \citet{Acciari-2014}. 
The frequencies/wavelengths
covered by the campaign are radio ($2.6-225$~GHz), near-infrared ($J$,
$H$ and $K$), optical ($B$, $V$, $g$, $R$ and $I$), UV
(\emph{Swift}/UVOT W1, W2 and M2), X-ray (0.3-195~keV), high-energy (HE) $\gamma$ rays
(0.1-400~GeV) and VHE ($0.08-5.0$~TeV).

Results on the broadband SED as well
as a detailed discussion of the SED modeling can be found in
\citetalias{Abdo-2011a}. It is the most detailed SED collected
simultaneously for Mrk\,421 during its typical activity state and the first time
where the high-energy component is completely covered by simultaneous
observations from the \emph{Fermi}-LAT and the VHE instrument
MAGIC. This allowed the characterization of the typical SED of
Mrk\,421 with unprecedented detail. In \citetalias{Abdo-2011a}, the SED
could be modeled reasonably well using either a one-zone synchrotron
self-Compton (SSC) model having two breaks in the electron spectrum,
or a hadronic (synchrotron proton blazar, SPB) model. In order to
distinguish between these two scenarios, one must look at the
multi-wavelength variability. One- and multizone SSC models predict a
positive correlation between X-ray and VHE flux
variations (e.g., \citealt{Graff-2008}), as they are produced by the
same electron population. In the SPB models of
\citetalias{Abdo-2011a}, a strict correlation between those two bands
is neither generally expected nor excluded, but can appear when
electrons and protons are accelerated together. Furthermore, the
one-zone SSC model of \citetalias{Abdo-2011a} predicts a correlation
between low-energy $\gamma$ rays from \emph{Fermi}-LAT with millimeter
(from SMA) and optical frequencies, something which would be hard to
incorporate in the SPB model, as the radiation is produced at
different sites.

In the following sections we will first characterize the flux
variability in all wavebands and then have a detailed look at the
cross-correlation functions between light curves of different bands,
primarily at X-rays vs. VHE and optical vs. X-rays, HE and VHE
correlations, but also at all other combinations as they might reveal
something interesting.

\section{Variability}\label{sec:variability}
\subsection{Light curves}
\begin{figure*}
  \includegraphics[height=1.\textheight]{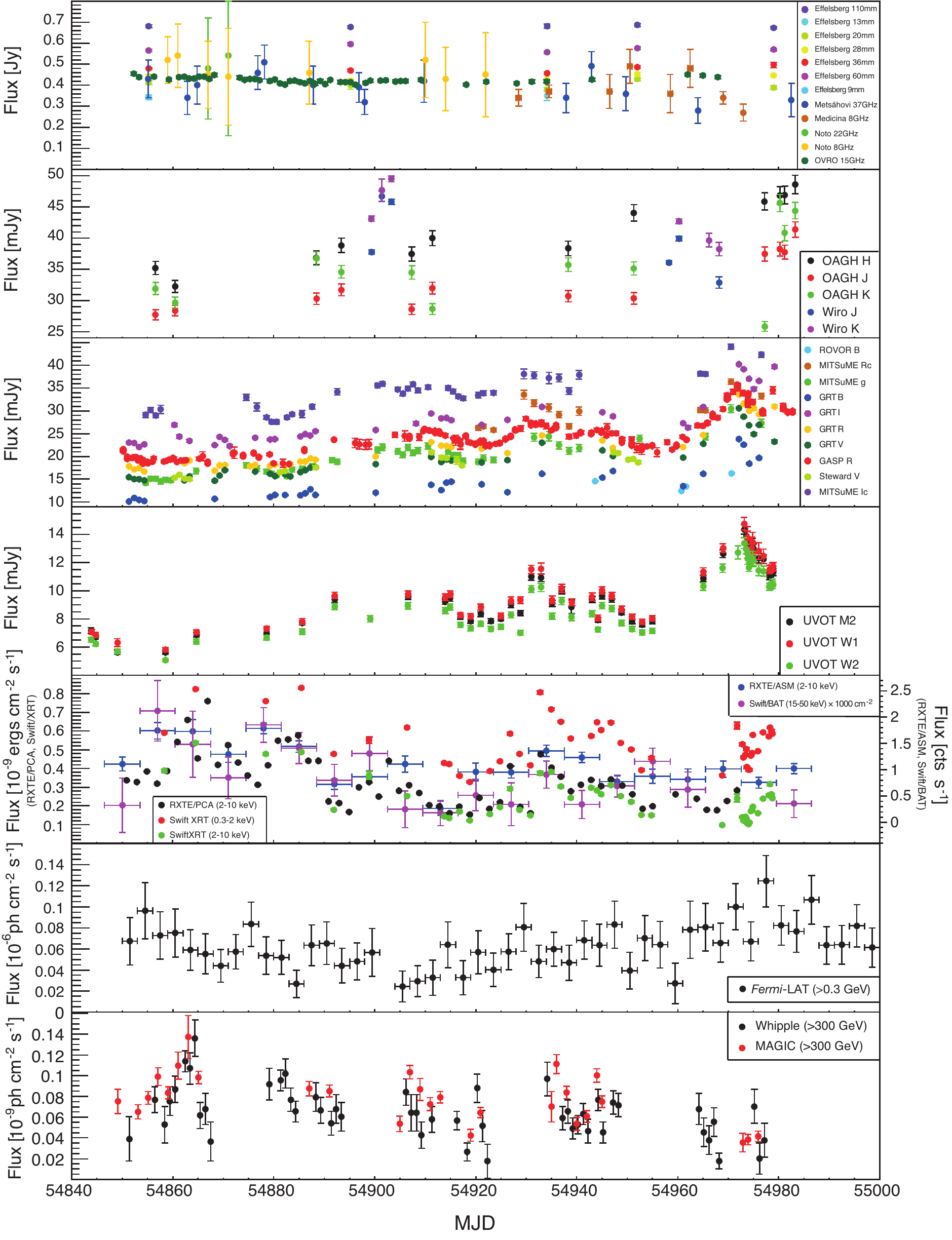}
  \caption{Light Curves of Mrk\,421 from radio to VHE from 2009 January 19 (MJD 54850) to 2009
   June 1st (MJD 54983). Vertical bars denote flux measurement errors, and the horizontal bars 
   denote the time bin widths into which some of the light curves are binned. The \emph{Fermi}-LAT
   photon fluxes are integrated over a three-day-long time interval. The Whipple 10-meter data (with an
   energy threshold of 400~GeV) were converted into fluxes above 300~GeV using a power-law
   spectrum with index of 2.5.}
\label{fig:lightcurves}
\end{figure*}

Figure \ref{fig:lightcurves} shows the light curves from radio to
VHE. No substantial (larger than a factor of $2$) flaring activity
happened during the campaign; however, some level of variability is
present in all energy bands. In the radio band the variability is
least pronounced. A significant level of variability is present in the
near-infrared (NIR), optical and UV accompanied by an overall increase
in flux with time. At X-ray, HE and VHE there is also
considerable variability, and only a small overall downward trend in
the overall X-ray and VHE flux with time is observed. The X-ray flux
variations are stronger on average than the variations in the other
wavebands, but still much weaker than the maximum values historically
registered for the X-ray and VHE bands \citep{Balokovic-2013,
Cortina-2013}.

\subsection{Fractional variability}\label{sec:Fvar} 

\begin{figure}
  \includegraphics[width=88mm]{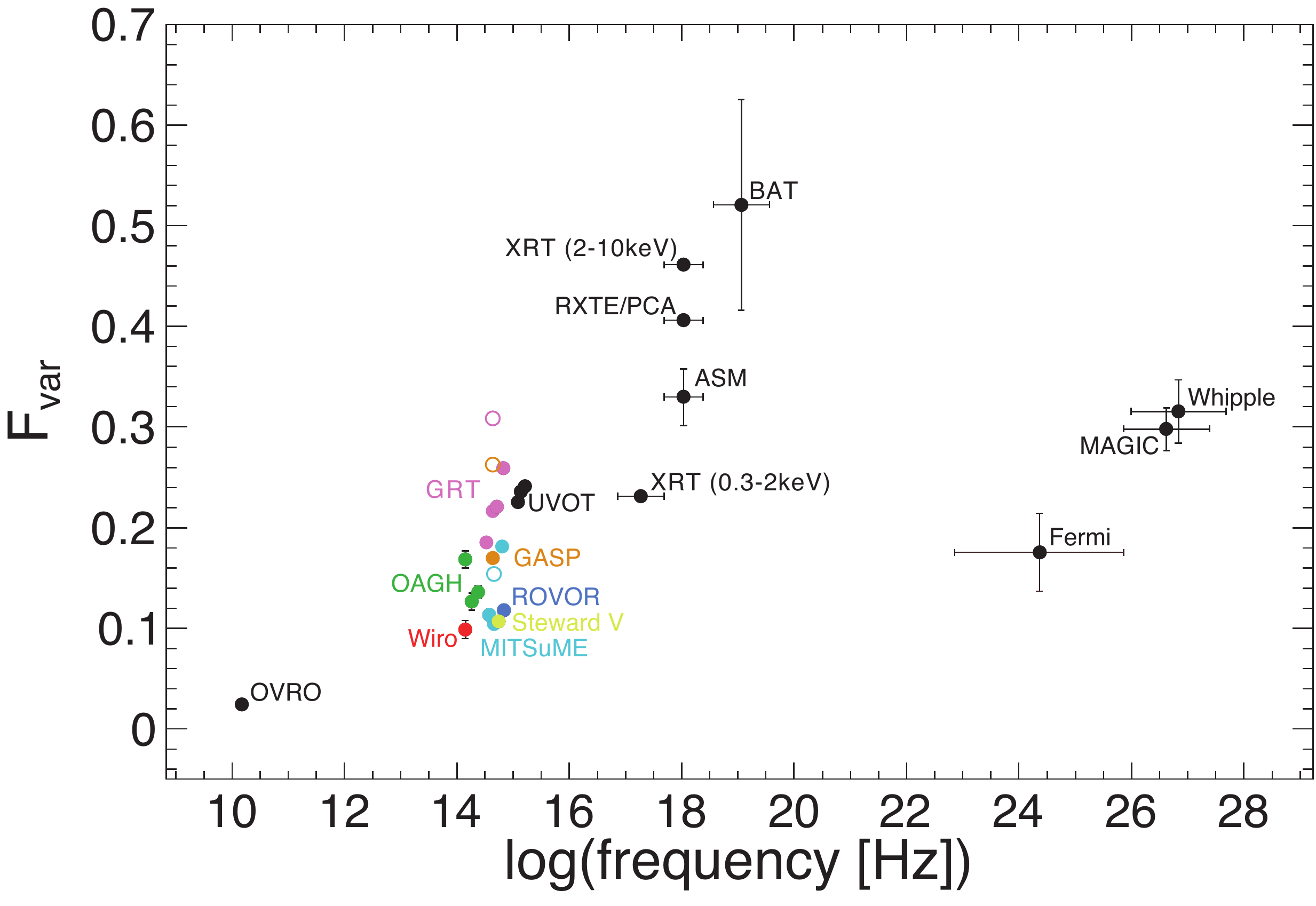}
  \caption{Fractional variability $F_{var}$ as a function of frequency. Open circles denote $F_{var}$
  values in $R$-band calculated with the host galaxy subtracted as prescribed in \citet{Nilsson-2007}}
\label{fig:Fvar}
\end{figure}

In order to quantify and characterize the variability at different
energy bands, we calculated the fractional variability 
\begin{equation}
F_{\mathrm{var}}=\sqrt{\frac{S^{2}-\left\langle\sigma^{2}_\mathrm{err}\right\rangle}{\left\langle x\right
\rangle^{2}}},
\end{equation}
i.e., the excess variance normalized by the flux, according to
\citet{Vaughan-2003}, where $S$ is the standard deviation of the $N$
flux measurements, $\left\langle\sigma^{2}_\mathrm{err}\right\rangle$ is the mean
squared error and $\left\langle x\right\rangle^{2}$ is the square of the average
photon flux. We estimate the uncertainty of $F_{\mathrm{var}}$
according to \citet{Poutanen-2008},
\begin{equation}
\Delta F_{\mathrm{var}}=\sqrt{F_{\mathrm{var}}^{2} + \mathrm{err}(\sigma_{\mathrm{NXS}}^{2})} - 
F_{\mathrm{var}},
\end{equation}
where $\mathrm{err}(\sigma_{\mathrm{NXS}}^{2})$ is given by equation 11 in \citet{Vaughan-2003}:
\begin{equation}
\mathrm{err}(\sigma_{\mathrm{NXS}}^{2})=\sqrt{\left(\sqrt{\frac{2}{N}}\frac{\left\langle\sigma^{2}_\mathrm{err}\right\rangle}{\left\langle x\right\rangle^{2}}\right)^{2} + \left(\sqrt{\frac{\left\langle\sigma^{2}_\mathrm{err}\right\rangle}{N}}\frac{2F_{\mathrm{var}}}{\left\langle x\right\rangle}\right)^{2}}.
\end{equation}
This prescription to calculate the uncertainties is more appropriate
than equation B2 in \citet{Vaughan-2003} for light curves that have an
error in the excess variance comparable to or larger than the
excess variance. This is, however, not the case for most light curves
in our sample, as $\Delta F_{\mathrm{var}}$ according to
\citet{Poutanen-2008} is less than $5$\% smaller compared to equation
B2 in \citet{Vaughan-2003}. For the \emph{Fermi}-LAT,
and \emph{Swift}/BAT light curves, the difference is
$\approx10$\%.

The $F_{\mathrm{var}}$ values and errors for the different energy
bands (instruments) are shown in Figure \ref{fig:Fvar}.  As already
noticed when looking at the light curves,  Mrk\,421 shows little variability 
in radio, and low but significant variability in all other wavebands with 
the largest variability in X-rays. We note that in the 2-10~keV band it is intrinsically
more variable than in the 0.3-2~keV band, a characteristic which has been recently
reported for Mrk\,421 during high X-ray and VHE activity \citep{Aleksic-2015b}

Because of the instrument sensitivity, the \emph{Fermi}-LAT, \emph{Swift}/BAT 
and RXTE/ASM data are binned into 3- and 7-day bins (instead of 1-day
bins) and therefore, the variability on smaller timescales is not
probed, so $F_{\mathrm{var}}$ might be underestimated. When rebinning
the RXTE/PCA light curve (sampled every $\sim2$~days) into 7-day bins,
$F_{\mathrm{var}}$ decreases by $\sim15$\% and agrees with
$F_{\mathrm{var}}$ for RXTE/ASM within the errors. The \emph{Swift}/XRT light curves are 
irregularly sampled.  There were measurements every $\sim7$~days during the early 
part of the campaign but there are also large gaps and a period of sub-daily observations. 
Rebinning the \emph{Swift}/XRT light curves into 1- and
7-day bins does not change $F_{\mathrm{var}}$ by more than a few
percent and all values agree within the errors.

The results reported in Figure \ref{fig:Fvar} are not affected by the
temporal binning or the uneven sampling of the light curves and hence
can be considered as characteristic of Mrk\,421 during the
multi-wavelength 2009 campaign (see Appendix A for details).

It is interesting to compare these results with the ones reported
recently for Mrk\,501 in \citet{Doert-2013} and \citet{Mrk501MW2008}, where the fractional
variability increases with energy and is largest at VHE, instead of
X-rays. The comparison of these observations indicates that there
are fundamental differences in the underlying particle populations,
environment, and/or processes producing the broadband radiation in
these two archetypical VHE blazars.  The higher X-ray variability in
Mrk\,421 might also be related to the higher synchrotron dominance
with respect to the one observed in Mrk\,501. According to the
broadband SEDs measured for Mrk\,501 and Mrk\,421 during the typical
(non-flaring) activity \citep{Abdo-2011a,Abdo-2011b}, $\nu
F_{\nu}^\mathrm{Sync_{peak}} >~ 2 \times \nu F_{\nu}^\mathrm{IC_{peak}}$ for Mrk\,501
and $\nu F_{\nu}^\mathrm{Sync_{peak}} >~ 4 \times \nu F_{\nu}^\mathrm{IC_{peak}}$ for
Mrk\,421. These SEDs were parametrized  within the one-zone SSC framework in
\citet{Abdo-2011a,Abdo-2011b}, using, for Mrk\,421, a magnetic field $B\sim2.5$ times 
higher than the one used for Mrk\,501 (38~mG vs. 15~mG), which naturally produces a synchrotron
bump that is relatively higher than the inverse-Compton bump. The higher magnetic field in
Mrk\,421 may also lead to a higher variability in the X-ray band
(with respect to the $\gamma$-ray bump) through a faster synchrotron
cooling of the high-energy electrons ($\tau_{\mathrm{cool-Sync}} \propto
1/B^2$).

\subsection{Evolution of the X-ray spectral shape with the X-ray flux}
\label{HarderWhenBrighter}
\begin{figure*}
\centering
  \includegraphics[width=17cm]{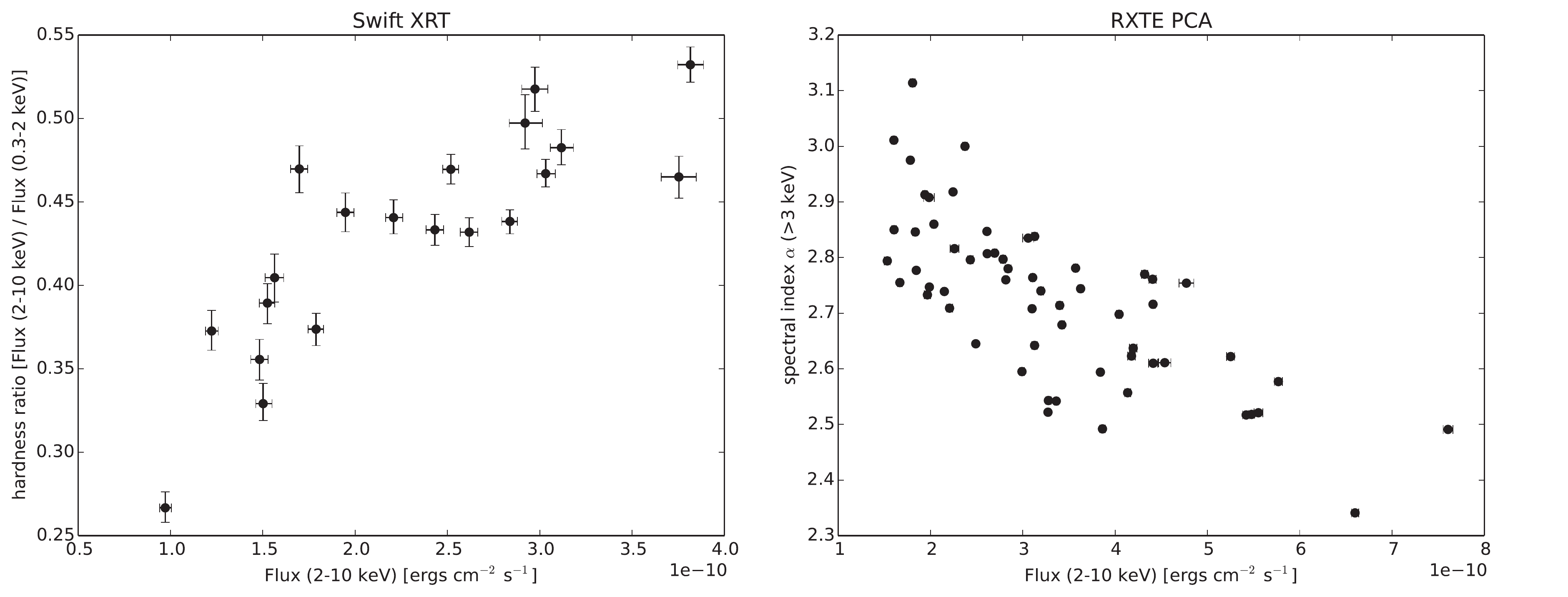}
  \caption{Left: X-ray hardness ratio for the
    \emph{Swift}/XRT bands 2-10~keV  and 0.3-2~keV vs. the X-ray flux in
    the 2-10~keV band. Right: Power-law index of RXTE/PCA spectra
    above 3~keV vs. the X-ray flux in the 2-10~keV band.}
\label{fig:alpha_flux_x}
\end{figure*}

Systematic variations of the X-ray spectral shape are a common phenomenon
during blazar flares \citep[e.g.][]{Fossati-2000}. A
harder-when-brighter behavior is quite typical during flares in blazars,
and this characteristic has already been observed in Mrk~421 \citep[e.g.][]{Tramacere-2009}.
Sometimes one can identify loops in the photon index vs. flux
diagram during the course of a flare, which could be related to the
dynamics of the system, as reported by \citet{Kirk-1999} or \citet{Rieger-2000}. 
Such behavior was also observed in Mrk~421 during a big flare in 1994 \citep{Takahashi-1996}.
Here we investigate whether these patterns exist in Mrk\,421 during its
typical (non-flaring) activity. 

The \emph{Swift}/XRT spectrum cannot be fit with a simple power-law because
this instrument covers the peak of the synchrotron bump, and hence the
X-ray spectrum in the 0.3-10 keV band is curved. We can
quantify the hardness of the \emph{Swift}/XRT spectra by using the ratio of the X-ray
fluxes in the bands 2-10 keV and 0.3-2 keV, and study its evolution
with respect to the X-ray flux in the 2-10 keV band. This is shown in the  left-hand
panel of Fig. \ref{fig:alpha_flux_x}. The RXTE/PCA spectra, which
cover the falling segment of the synchrotron bump (when the source is
not flaring), can be fit with a simple power-law function, and hence
here we can report the spectral slope vs. the X-ray flux in the
2-10 keV band. This is shown in the in the right-hand panel of
Fig. \ref{fig:alpha_flux_x}. Both \emph{Swift} and RXTE show clearly
that the X-ray
spectra harden when the X-ray emission increases, and hence we
can confirm that the harder-when-brighter behavior also occurs
when the source is not flaring. We also investigated the temporal
evolution of the plots shown in Fig.~\ref{fig:alpha_flux_x}, looking for loop patterns 
in the spectral shape-flux plots (clockwise
or counter-clockwise) but we did not find any.

\subsection{Power density spectrum}\label{sec:psd} 
Another way to characterize the variability of a given source is the
power spectral density (PSD). The PSD quantifies the variability
amplitude as a function of Fourier frequency (or timescale) of the
variations. The derivation of the PSD is based on the discrete Fourier
transform of the light curve under consideration. For blazars, the
shape of the PSD is usually a power-law $P_{\nu}\propto\nu^{-\alpha}$
with spectral index $\alpha$ between $1$ and $2$ 
\citep{Abdo-2010,Chatterjee-2012}, i.e., there is larger variability
at smaller frequencies / longer timescales. This is generally referred
to as ``red noise''. Other features in the PSD, such as breaks or
peaks, indicate characteristic timescales or (quasi)\-periodic signals.

Calculating the PSD via the discrete Fourier transform is
straightforward for light curves that are frequently and regularly
sampled over a long period of time. However, in reality, observation
time is usually limited and often interrupted by bad weather, object
visibility and technical issues, i.e., we are normally dealing with
unevenly sampled light curves of limited length that may have large
gaps, and this has serious effects on the measured PSD. If the
light curve is discretely sampled instead of continuous (which is
usually the case as we are dealing with discrete observations or
values that are binned over a certain time period), its Fourier
transform is convolved with a windowing function, which becomes
very complicated when a light curve has an uneven sampling and gaps
\citep{Merrifield-1994}. In addition, light curves of a finite length
are affected by red-noise leak, i.e., variability below the smallest
frequency (or largest timescale) probed. This variability power leaks
into the observed frequency band and changes the observed PSD
shape. This effect can manifest as a rise/fall trend throughout the entire 
time interval of the light curve. Likewise,
aliasing, i.e., variability power from frequencies larger than the
Nyquist frequency, affects the variability in the observed frequency
range. The effect of sampling on the study of periodicities will be
discussed in Section \ref{sec:periodicities}.

These effects of the sampling pattern on the PSD can be avoided by
using the simulation-based approach of \citet{Uttley-2002} (PSRESP) to 
derive the intrinsic PSD of a light curve and its associated
uncertainties. We applied this Monte Carlo fitting technique following
the prescription given in the appendix of \citet{Chatterjee-2008} to
all light curves with $\sim30$ or more flux measurements.

First we generated a large set of simulated light curves using the
method of \citet{Timmer-1995}. In order to accommodate the problems
introduced by the sampling of the light curve, the simulated
light curves were about 100 times longer than the measured light curve
and then clipped to the required length. This way, they suffer from
red-noise leak in the same way as the measured light curve. The
simulated light curves also had a much finer sampling than the measured
light curve and were then binned to the required binning to include the
aliasing effect. Finally, the simulated light curves were
resampled with the observed sampling function, so that the windowing
function is the same for the measured and the simulated
light curves. Poisson noise was added to each simulated light curve to
account for observational noise. As a model for the underlying PSD of
the simulated light curves we assumed a power-law shape and
varied the power-law index $\alpha$ in the range $1.0$ to $2.5$ in
steps of $0.1$. We generated $1000$ simulated light curves per $\alpha$
value and per measured light curve. We then calculated the PSD for each 
measured light curve and the 1000 simulated light curves of each model, 
taking as the PSD the modulus squared of the mean subtracted light curves'
discrete Fourier transform between the minimum frequency
$\nu_{min}=1/T$ and the Nyquist frequency $\nu_{Ny}=N/2T$. $T$ is the
duration of the light curve. The frequency range covered by our data is
approximately $10^{-7} - \gtrsim10^{-5.7}$~s$^{-1}$ (corresponding to
$\approx1/120$~days$^{-1}$ - $\approx1/6$~days$^{-1}$), differing
somewhat between the light curves depending on the individual length
and binning. The light curves were binned into $2 - 7$-day bins,
depending on the light curve characteristics. The goodness-of-fit of each 
PSD model was determined according to the recipe given in the appendix of 
\citet{Chatterjee-2008}: The observed $\chi^2$ function 
\begin{equation}
\chi^2_\mathrm{obs}=\sum^{\nu_\mathrm{max}}_{\nu=\nu_\mathrm{min}}
\frac{\left(\mathrm{PSD}_\mathrm{obs}-\overline{\mathrm{PSD}}_\mathrm{sim}\right)^2}
{\left(\Delta\mathrm{PSD}_\mathrm{sim}\right)^2} 
\end{equation}
from the observed PSD$_\mathrm{obs}$, the average of the 1000 PSDs from 
simulated light curves $\overline{\mathrm{PSD}}_\mathrm{sim}$, and the standard 
deviation $\Delta\mathrm{PSD}_\mathrm{sim}$ was compared 
to the simulated $\chi^2$ distribution 
\begin{equation}
\chi^2_{\mathrm{dist}, i}=\sum^{\nu_\mathrm{max}}_{\nu=\nu_\mathrm{min}}
\frac{\left(\mathrm{PSD}_{\mathrm{sim}, i}-\overline{\mathrm{PSD}}_\mathrm{sim}\right)^2}
{\left(\Delta\mathrm{PSD}_\mathrm{sim}\right)^2} 
\end{equation}
calculated from each of the 1000 individual PSDs from simulated light curves 
$\mathrm{PSD}_{\mathrm{sim}, i}$. The success fraction (SuF) is then the 
fraction of $\chi^2_{\mathrm{dist}, i}$ larger than $\chi^2_\mathrm{obs}$.

\begin{figure*}
  \includegraphics[width=17cm]{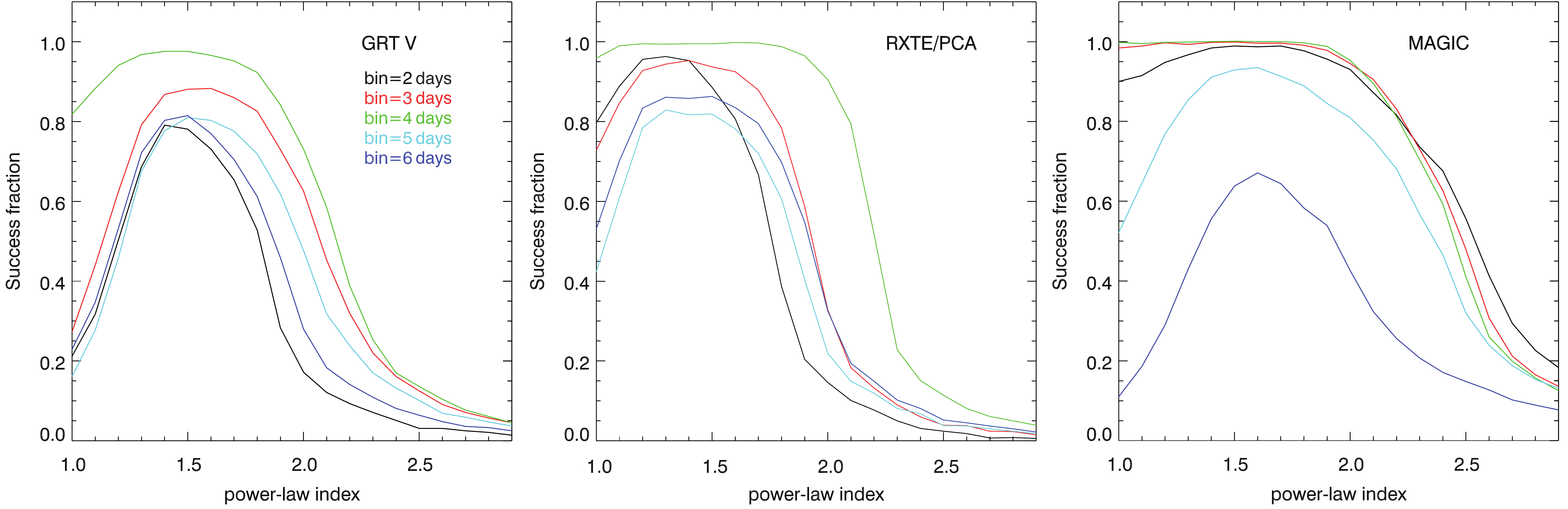}
  \caption{Success fraction as a function of power-law index $\alpha$
    of the PSD for three selected light curves (GRT $V$ band,
    RXTE/PCA and MAGIC) and a range of light curve binnings
    ($2 - 6$~days) to illustrate the effect of the binning. The
    location of the maximum does not change significantly with the
    binning, but there is significant variation in shape, width and
    amplitude.}
\label{fig:success_fraction}
\end{figure*}

Figure \ref{fig:success_fraction} shows the SuFs for selected measured
light curves and Table \ref{tab:psdslope} gives the best-fit
power-law indices $\alpha$ and their uncertainties, calculated as the
half width at half maximum (HWHM) of the SuF vs. $\alpha$ curve.

We tried a range of binnings and found that the location of the
maximum does not vary systematically with the binning
(Figure \ref{fig:success_fraction}). The uncertainties, however, depend
strongly on the light curve binning and on the logarithmic binning of
the PSD \citep{Papadakis-1993}. We used light curve bin sizes of a few
days (between $2$ and $6$) and a factor of $1.2$ or $1.3$ by which the
logarithmically spaced frequency bins are separated, in order to
reduce the scatter in the PSD points. Sometimes the flux measurements
have large uncertainties (mean error non-negligible compared to the
variance of the light curve as, e.g., in the case of \emph{Fermi}-LAT)
or the light curve has large gaps and/or relatively few data points
(e.g., MAGIC). In these cases $\alpha$ is mostly unconstrained. Large
gaps or very uneven binning may result in large SuF differences (e.g.,
MAGIC), or even in changes of the overall shape (e.g., OVRO). In these
cases there is no good way to bin the data and obtain a reliable
$\alpha$. If the binning is too small, large gaps are filled with
interpolated (probably unrealistic) data in order to calculate the
discrete Fourier transform. This results in unwanted changes in the
reconstructed PSD and thus in unreliable $\alpha$ values. If the
binning is so large that it accomodates also the large gaps such that
not too much interpolation is necessary, there are too few data points
left and the covered frequency interval becomes too small for a
reasonable PSD fit. Unfortunately, the light curves are too short to
make analyses based on contiguous parts of a light curve between large
gaps.

The maximum of the SuF vs. $\alpha$ curve is generally $\gtrsim0.8$
for all light curves, but we note that for certain binnings the maximum
SuF can be significantly lower, or saturate at 1.0. Thus a
power-law seems to be a reasonably good fit for all light curves, but
as explained above, the fit is restricted by the limited frequency
range and by the light curve sampling or gaps. In Table
\ref{tab:psdslope} we give the $\alpha$ values where the SuF has a
maximum, i.e., the best-fitting power-law indices. As uncertainties we
mention only the median HWHM of the distribution of HWHM from
different light curve binnings between $2$ and $6$~days, but please note that
this value itself has an uncertainty. In many cases it is not clear
why we should prefer one binning over another, so a certain range in
SuF shapes is possible. It should be pointed out that these uncertainties in
deriving the width of the SuF vs. $\alpha$ curve do not affect the
analysis in the following sections, as we always use the best-fitting
$\alpha$, which does not vary with the binning.

\begin{table}
\caption{Power spectral density (PSD) index $\alpha$ and half-width at half-maximum of the success 
fraction (SuF) for light curves with more than 30 flux measurements.}       
\label{tab:psdslope}   
\centering   
\begin{tabular}{l c c}  
\hline\hline  
Instrument & $\alpha$\tablefootmark{a} & HWHM \\ 
\hline  
OVRO & 2.0 & $1.2 - 2.3$\tablefootmark{b} \\
GRT I & 1.5 & 0.7\\
MITSuME Ic & 1.6 & 0.7\\
GASP & 1.9 & 0.5 \\
GRT R & 1.4 & 0.8 \\
GRT V & 1.5 & 0.5\\
MITSuME g & 1.5 & 0.6 \\
GRT B & 1.4 & 1.0 \\
UVOT W1 & 1.4 & 0.5 \\
\emph{Swift}/XRT (0.3-2 keV) & 1.5 & 0.6 \\
\emph{Swift}/XRT (2-10 keV) & 1.4 & 0.6 \\
RXTE/PCA & 1.4 & 0.6 \\
\emph{Fermi}-LAT &  & $1.0 - 2.8$\tablefootmark{c}\\
MAGIC & 1.6 & 0.9 \\
Whipple & 1.3 & 0.6 \\
\hline  
\end{tabular}
\tablefoot{ \\
\tablefoottext{a}{slope of a power-law $P(\nu)\propto\nu^{-\alpha}}$} \\
\tablefoottext{b}{The SuF vs. $\alpha$ curve is asymmetric, thus the PSD 
of OVRO is not very well constrained. This might be due to the very low variability 
in radio compared to the other wavelengths.} \\
\tablefoottext{c}{The PSD of \emph{Fermi}-LAT is largely 
unconstrained because of the large flux measurement errors. The SuF is 
approximately constant and high ($\textgreater0.8$) over a large 
$\alpha$ range with no clear maximum.}
\end{table}

The best-fitting PSD models for most light curves are found to be
power-laws with indices $\sim1.3 - 1.6$. There are no big
differences between the SuF vs. $\alpha$ curves of different
instruments. The only exceptions are GASP $R$ band and OVRO, where the
shape is different and $\alpha$ is higher (though still in agreement
with the other $\alpha$ values within the uncertainties). A possible
explanation might be the small flux error bars and the dense sampling
(several values per night, i.e., larger frequency range compared to all
other light curves). Without logarithmic binning and with a light curve
binning of 1 or 2 days there is a clear maximum at $\alpha=1.8$.

There is no evidence of a break in the (relatively short) frequency range
covered by the multi-wavelength data. Simulated light curves with an
underlying broken power-law PSD did not improve the success
fraction. The X-ray PSD power-law indices are similar to the ones reported
in \citet{Kataoka-2001} for the same frequency range. The PSD shape
and $\alpha$ are consistent with what was found for blazars by other
authors (\citealt{Chatterjee-2008}: X-rays; \citealt{Abdo-2010}:
HE $\gamma$ rays).

\section{Cross-correlations}\label{sec:correlations}

We use the discrete cross-correlation function (DCF) method of
\citet{Edelson-1988} to quantify the correlation of the flux
variations between all possible light curve pairs, as long as the
light curve has more than 30 flux measurements, i.e., we use the same
set of light curves as in Section \ref{sec:psd}. This way we can assess
correlations between VHE, HE $\gamma$ rays, X-ray, UV, optical and some
radio frequencies. 

As a cross-check, we calculate for each light curve pair also the
$z$-transformed cross-correlation function (ZDCF;
\citealp{Alexander-1997}). The ZDCF is based on the DCF, but the bin
widths of the ZDCF are chosen such that the number of points is the
same for each bin, i.e., they are different-sized as opposed to the
DCF, where all bins have the same size. In addition, Fisher's
$z$-transform is applied to the cross-correlation coefficients.
According to \citet{Larsson-2012}, the ZDCF is more robust than DCF for 
undersampled (w.r.t. the flux variations) light curves. However, for well 
sampled light curves, the ZDCF has been shown to be consistent with the 
DCF \citep[e.g.,][]{Dietrich-1998, Smith-2007}. For this study we used 
mostly the DCF, as the temporal bin is fixed, so that it also allows
us to trivially
compare and even combine results from  different pairs of instruments. 
Therefore, we used the ZDCF for verification purposes only.

Two different approaches are used to determine the uncertainties of
the DCF. The easiest and fastest way is to simply use the errors given
in \citet{Edelson-1988}. However, as discussed in \citet{Uttley-2003},
these are not appropriate for determining the significance of the DCF
when the individual light curve data points are correlated red-noise
data. Depending on the PSD and the sampling pattern, the significance
as calculated by \citet{Edelson-1988} might be overestimated. To get a
better estimate on the real significance of the correlation peaks we
used a Monte Carlo approach, following the descripion of
\citet{Arevalo-2009}. The Monte Carlo technique is described in detail
in Section \ref{sec:corr_vhe_x}.

We used a binning of six days for all DCFs because this way different
DCFs can be easily compared or, if needed, combined. To make sure that
we do not miss correlations or time lags, we always tried a range of
binnings, depending on the sampling of the involved light curves.

As the following paragraphs will show, significant correlations are
only found between X-rays and VHE. In addition, X-rays and optical
light curves seem to follow opposite trends.

\subsection{VHE --- X-rays}\label{sec:corr_vhe_x}

\begin{figure*}
  \includegraphics[width=17cm]{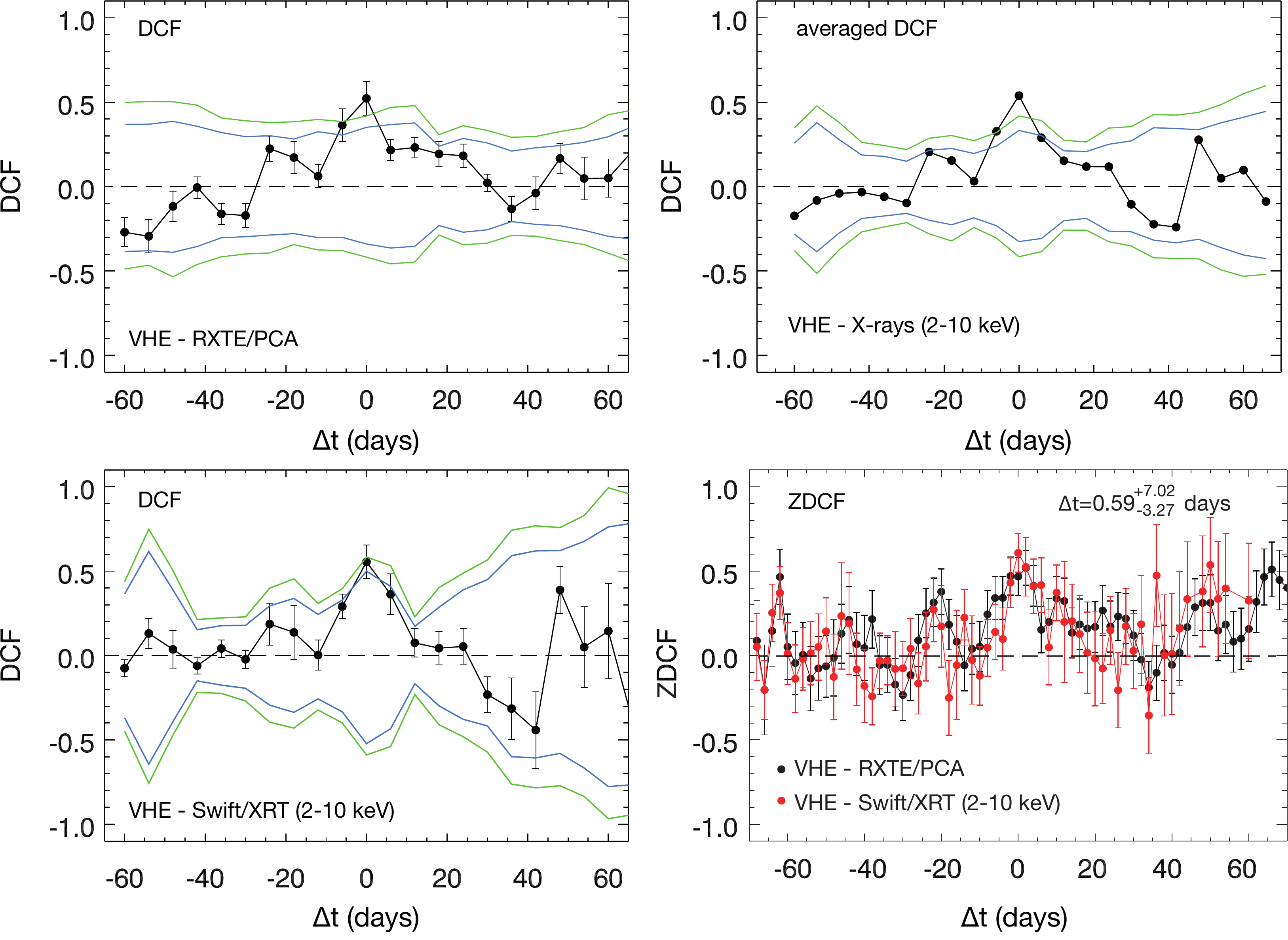}
  \caption{The DCF of the combined Whipple and MAGIC (VHE)
    light curve, correlated with the RXTE/PCA and the \emph{Swift}/XRT
    (2-10~keV) light curves are shown in the upper left and lower left
    panels. The black error bars represent the uncertainties as
    derived from \citet{Edelson-1988}. The green lines represent the
    1\% and 99\% extremes of the DCF distribution of simulated
    RXTE/PCA light curves when correlated with the measured VHE
    light curve. The blue lines represent the 5\% and 95\%
    extremes. The upper right panel shows the average of the
    VHE--RXTE/PCA and VHE--\emph{Swift}/XRT (2-10~keV) DCFs, with the
    corresponding confidence intervals derived from averaging the DCFs
    of the simulated light curves. See text for details in the
    calculation of the average DCFs and contours. The lower right
    panel shows the $z$-transformed DCFs.}
\label{fig:dcf_vhex}
\end{figure*}

The correlation of the flux variations between the VHE (MAGIC,
Whipple) and X-ray (RXTE/PCA and \emph{Swift}/XRT 2-10keV) bands is
shown in Figure \ref{fig:dcf_vhex}. The correlations peak at time lag
$\Delta t=0$ and appear to be strongly significant
($\textgreater5\mathrm{\sigma}$), when considering only the errors
calculated according to \citet{Edelson-1988} (black error
bars). However, as mentioned above, the error bars calculated using \citet{Edelson-1988} can
overestimate the real significance of the correlation.  To get a better estimate on the real
significance of the correlation peaks we use the following Monte Carlo
approach.

For each X-ray light curve we created a set of $1000$ simulated
light curves in the same way as in Section \ref{sec:psd}, using a
power-law with the best-fitting slope as determined in Section
\ref{sec:psd} from the PSRESP method (see Table
\ref{tab:psdslope}). The X-ray flux was sampled more often and with
smaller statistical errors than the VHE flux, and hence, in order to
ascertain the confidence levels in the DCF calculation, it is
reasonable to use simulated RXTE/PCA and \emph{Swift}/XRT ($2-10$~keV)
light curves instead of the VHE light curves. We cross-correlated each
of the $1000$ simulated X-ray light curves with the observed MAGIC and
Whipple light curves. A power-law spectrum with index $2.5$ \citep{Hillas-1998} was used to
normalize the integral flux of the Whipple $10$~m data (with an energy
threshold of $\sim400$~GeV) to an integral flux above $300$~GeV in
order to provide a comparison with MAGIC.  We therefore also
cross-correlated the original and the simulated X-ray light curves with
the combined Whipple+MAGIC light curve. From the distribution of $1000$
DCFs (i.e., the cross-correlations of the simulated X-ray light curves
with the real VHE light curves) we then calculated the $95$ and $99\%$
confidence limits, and show them in Figure \ref{fig:dcf_vhex}. For
each combination of RXTE/PCA and \emph{Swift}/XRT (2-10~keV) with MAGIC,
Whipple and Whipple+MAGIC the DCF shows a peak at time lag $\Delta
t=0$ with a probability larger than $99$\%. There are no other peaks or
dips in the DCF between VHE and X-rays that appear significant. A
positive correlation between X-rays and VHE has been reported
many times during flaring activity \citep[e.g., ][]{Fossati-2008}, but
has never been observed for Mrk\,421 in a non-flaring state. Our
simulations show that the real significance of the correlation is
$3-4\mathrm{\sigma}$, indeed confirming that the error bars calculated
using \citet{Edelson-1988} slightly overestimate the significance of the
correlation.

We average the DCFs, which has the advantage that spurious features
are smoothed out while features present in all DCFs (i.e., those
features that are real) are strengthened. This is particularly useful
when having many possible combinations and/or marginally significant
features like the ones that will be reported in Section
\ref{sec:dcf_xopt}. The DCFs were averaged in the following way: For a
number of $q+1$ real light curves $A$, $B_1$, $\ldots$, $B_q$ we first
calculate all $q$ correlation functions DCF($AB_1$), $\ldots$,
DCF($AB_q$) using a binning of $6$~days. Then we calculate the average
DCF
\begin{equation}
\overline{\mathrm{DCF}}=\frac{1}{q}\sum\limits_{i=1}^{q}\mathrm{DCF}(AB_i).
\end{equation}
There is no prescription to combine the uncertainties derived from
\citet{Edelson-1988} for several DCFs, thus no error bars are shown in the 
upper right panel of Figure \ref{fig:dcf_vhex}. For the determination of the
averaged confidence limits we correlate the simulated light curves
$a_1$, $\ldots$, $a_n$ or $b_{\tilde{q},1}$, $\ldots$,
$b_{\tilde{q},n}$ ($n=1000$, $\tilde{q}\in1,\ldots,q$) with the
original light curve $B_{\tilde{q}}$ or $A$ using either
\begin{equation}\label{eqn:2}
\overline{\mathrm{DCF}_{\mathrm{sim},j}}=\frac{1}{q}\sum\limits_{i=1}^{q}\mathrm{DCF}(a_jB_i) \hspace{1cm}\forall j=1,\ldots,n
\end{equation}
or
\begin{equation}\label{eqn:3}
\overline{\mathrm{DCF}_{\mathrm{sim},j}}=\frac{1}{q}\sum\limits_{i=1}^{q}\mathrm{DCF}(Ab_{i,j}) \hspace{1cm}\forall j=1,\ldots,n.
\end{equation}
From this distribution of $n$ averaged correlation functions
$\overline{\mathrm{DCF}_{\mathrm{sim},j}}$ we then compute the $95$
and $99$\% confidence limits. Whether we use equation \ref{eqn:2} or
equation \ref{eqn:3} depends on the sampling of the light curves. The
sampling and statistical uncertainties of the X-ray light curves are
much better than the sampling of the VHE light curves. Therefore the
PSD could be better constrained in the X-ray case and hence the
confidence limits obtained from correlating the original VHE with
simulated X-ray light curves are more reliable than the confidence limits
obtained from correlating the original X-ray with simulated VHE
light curves. Thus here the light curve $A$ is Whipple+MAGIC, the
light curves $B_i$ are RXTE/PCA and \emph{Swift}/XRT (2-10~keV) and we use
equation \ref{eqn:3}.

The upper right panel of Figure \ref{fig:dcf_vhex} shows the DCF averaged over
the two combinations Whipple+MAGIC -- RXTE/PCA and Whipple+MAGIC --
\emph{Swift}/XRT (2-10~keV) and the corresponding confidence limits. There is
a clear correlation at time lag $\Delta t=0$ with a high confidence
$>99$\%.

The ZDCFs between the VHE and the X-ray light curves show the same
behavior as the corresponding DCFs. However, as the binning is
different for each ZDCF (ranging from sub-day scales around time lags
$\Delta t\approx 0$~days to a few days at time lags $\Delta t\approx
60$~days), it is not possible to combine them as we did with the
DCFs. Rebinning the ZDCFs to even 6-day bins, averaging them
subsequently and using the simulated light curves to assess the
uncertainties yields almost identical results to the averaged DCF.

Both VHE and X-ray light curves show a weak negative trend with
time. To make sure that this trend is not responsible
for the correlation, we calculated the (z)DCFs also for the detrended
light curves. The difference is marginal. In addition, when comparing X-ray
and VHE light curves, one can see that the light curve features nicely
agree, i.e., the long-term trend has only a minor contribution to the
correlation peak, which is driven by shorter timescale variability.

\begin{figure}
  \includegraphics[width=87mm]{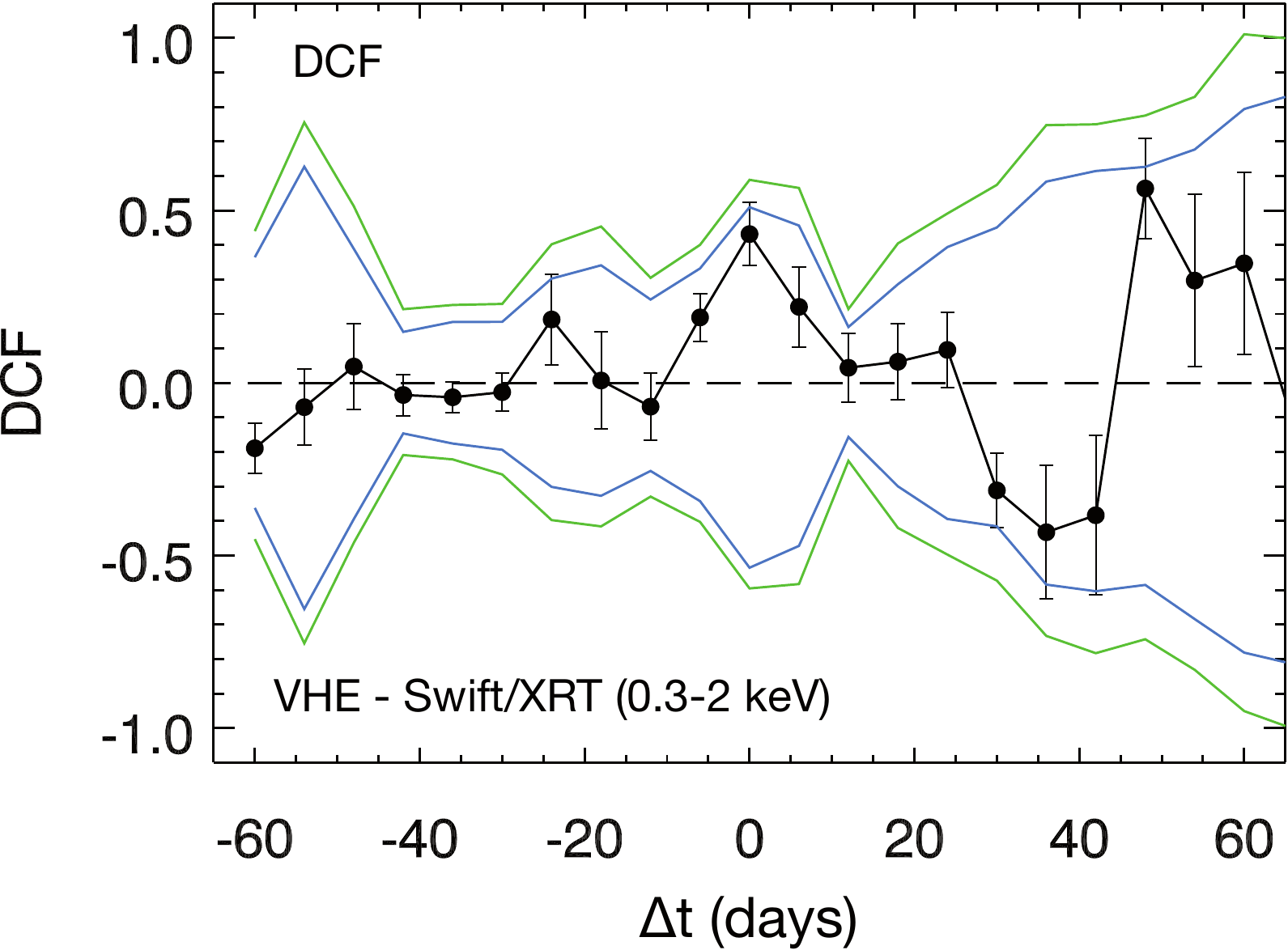}
  \caption{DCF of the combined Whipple and MAGIC (``VHE'') light curve,
    correlated with the \emph{Swift}/XRT (0.3-2~keV) light curve. The black
    error bars represent the uncertainties as derived from
    \citet{Edelson-1988}. The green lines represent the 1\% and 99\%
    extremes of the DCF distribution of simulated \emph{Swift}/XRT
    light curves when correlated with the measured VHE light curve. The
    blue lines represent the 5\% and 95\% extremes. }
\label{fig:dcf_vhe_xrta}
\end{figure}

Figure \ref{fig:dcf_vhe_xrta} shows the DCF of the combined Whipple and
MAGIC light curve, correlated with the \emph{Swift}/XRT (0.3-2~keV)
light curve. Although the DCF has a peak at time lag $\Delta
t\approx0$~days which seems to be significant with a confidence level
of around $5\mathrm{\sigma}$ when considering the \citet{Edelson-1988} errors,
the simulations show that this level of correlation is not significant
(only $\approx1.5\mathrm{\sigma}$).

\subsection{HE $\gamma$ rays --- UV/optical} \label{sec:dcf_gammaopt}
\begin{figure}
  \includegraphics[width=88mm]{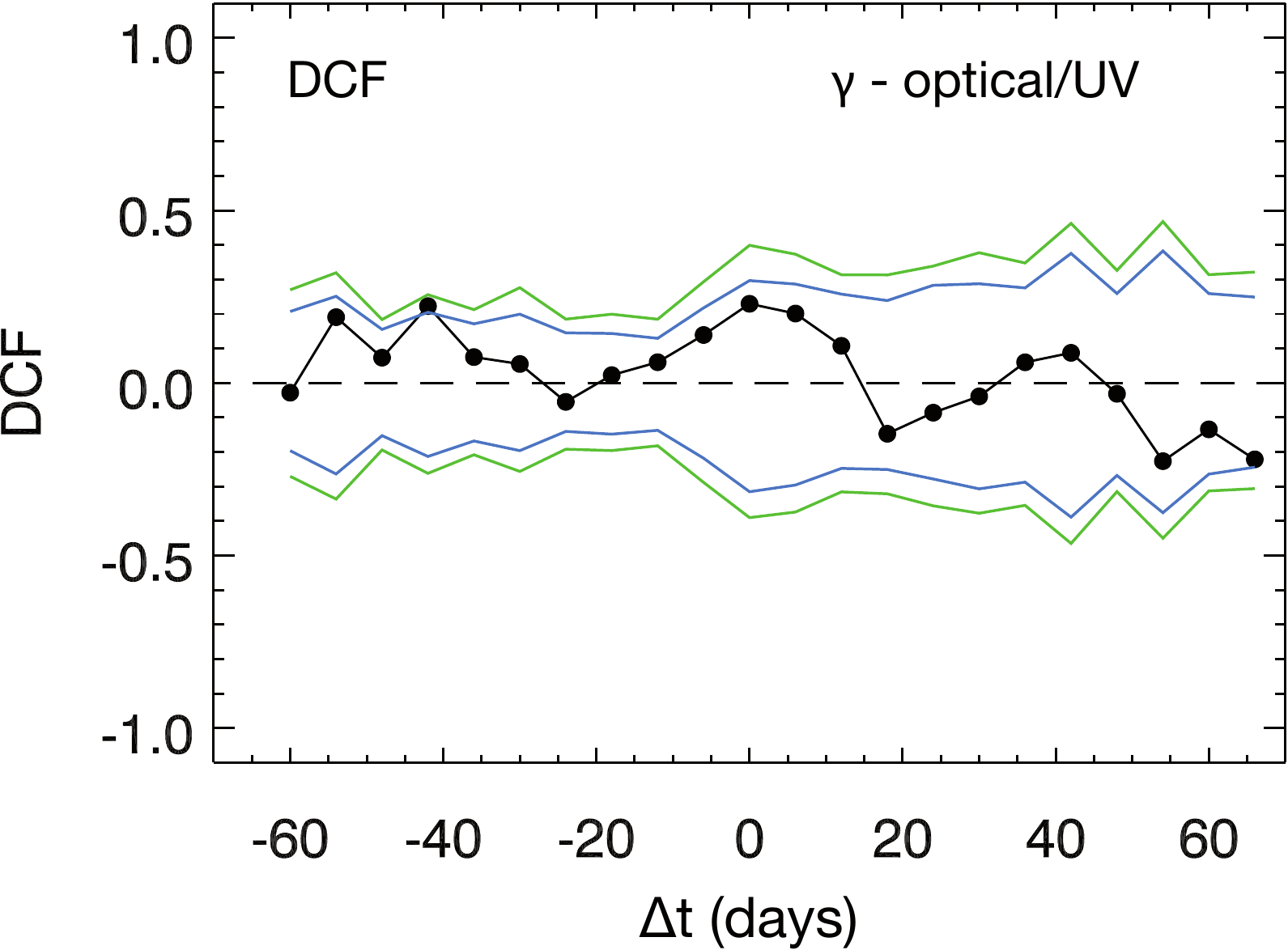}
  \caption{The average DCF of the \emph{Fermi}-LAT HE $\gamma$-ray
    light curve,
    correlated with the optical and UV light curves (GASP $R$-band, GRT $BVRI$, MITSuME g, 
    MITSuME Ic and UVOT W1), is shown in black. The green
    lines represent the 1\% and 99\% extremes of the likewise
    averaged DCF distribution of simulated optical/UV light curves when
    correlated with the real \emph{Fermi}-LAT light curve. The blue
    lines represent the 5\% and 95\% extremes.}
\label{fig:dcf_gammaopt}
\end{figure}

Figure \ref{fig:dcf_gammaopt} shows the combined DCF of the
\emph{Fermi}-LAT light curve, correlated with optical and UV
light curves (using simulated optical and UV light curves to estimate
the uncertainties)\footnote{In Section \ref{sec:psd} we showed that it is
  not possible to constrain the PSD of the \emph{Fermi}-LAT light curve
  because of the large error bars of the light curve data points.}. There
is a peak in the DCF at a time lag $\Delta t = 0$~days, but it is not
significant. The uncombined DCFs also show a small peak with a
significance $\textless3\mathrm{\sigma}$ or no peak at all when using
the \citet{Edelson-1988} uncertainties. The significance of the small
peaks is even lower ($\textless2\mathrm{\sigma}$) when using the
uncertainties from simulated data. Larger light curve binsizes would
reduce the errors, but also lead to significantly fewer data points, a
reduced PSD frequency range, and increased DCF bin sizes. Given the
small time window (4.5 month long campaign) under consideration, we
cannot improve the DCF quality by rebinning.

\subsection{X-rays --- UV/optical} \label{sec:dcf_xopt}
\begin{figure*}
  \includegraphics[width=17cm]{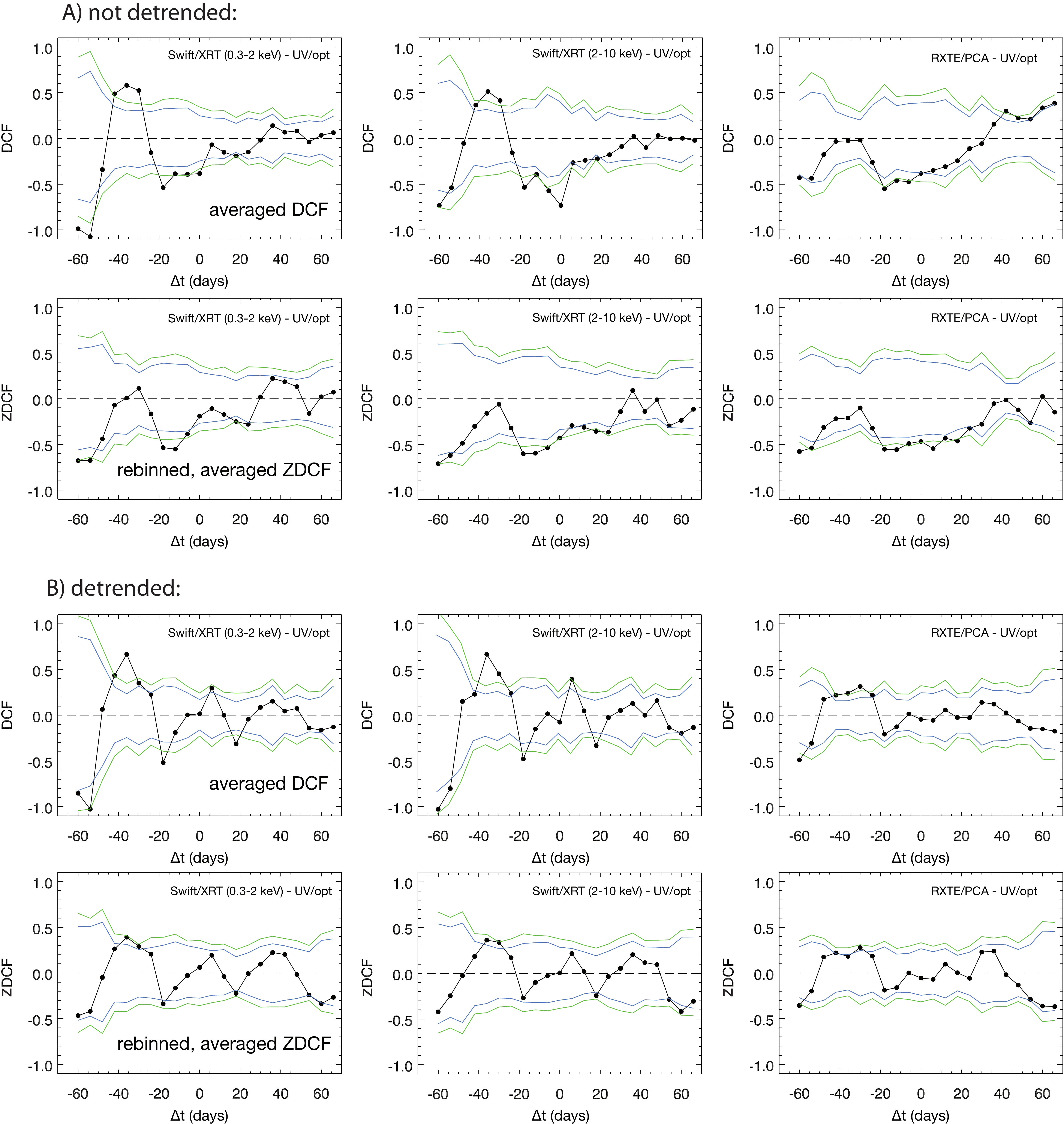}
  \caption{(A) The DCFs of each X-ray light curve (\emph{Swift}/XRT
    (0.3-2~keV), \emph{Swift}/XRT (2-10~keV) and RXTE/PCA (2-10~keV)), correlated with
    several optical and UV light curves, are averaged over all optical
    to UV bands and shown in black in the upper panel. The green lines
    represent the 1\% and 99\% extremes of the likewise averaged DCF
    distribution of simulated optical/UV light curves when correlated
    with the observed X-ray light curve. The blue lines represent the
    5\% and 95\% extremes. The lower panel shows the z-transformed
    DCFs, which were, for the purpose of direct comparison with the
    DCF, rebinned to the same binning as the DCFs and averaged in the
    same way. (B) shows the same as (A), but all light curves have been
    detrended (as described in \ref{sec:dcf_xopt}) before correlation.}
\label{fig:dcf_xopt}
\end{figure*}

We also searched for correlations between the X-ray and UV/optical
bands. As done in the previous subsections, we calculated the DCFs for
all possible combinations between RXTE/PCA and \emph{Swift}/XRT with
\emph{Swift}/UVOT, GASP R band, GRT BVRI, MITSuME g and MITSuME Ic. The
results are shown in panel A of Figure \ref{fig:dcf_xopt}. One
feature that is common in almost all DCFs is an anti-correlation at a
time lag $\Delta t\approx-20$ to $-10$~days, i.e., optical/UV variations lead
X-ray variations by $\sim15$~days. This feature is significant above
$99$\% and is confirmed by the ZDCF. However, it is not immediately
clear what might cause this anti-correlation. The first thing that
becomes apparent when looking at the light curves is the long-term
trend. The UV/optical light curves show a strong positive trend, while
the X-ray light curves show a slight negative trend. Therefore it is
not surprising that the DCFs show an overall anti-correlation spread
over a large range of time lags. However, this characteristic cannot
explain the above-mentioned (anti-)correlations with time lags of
$10$-$20$~days.

\begin{figure*}
  \includegraphics[width=17cm]{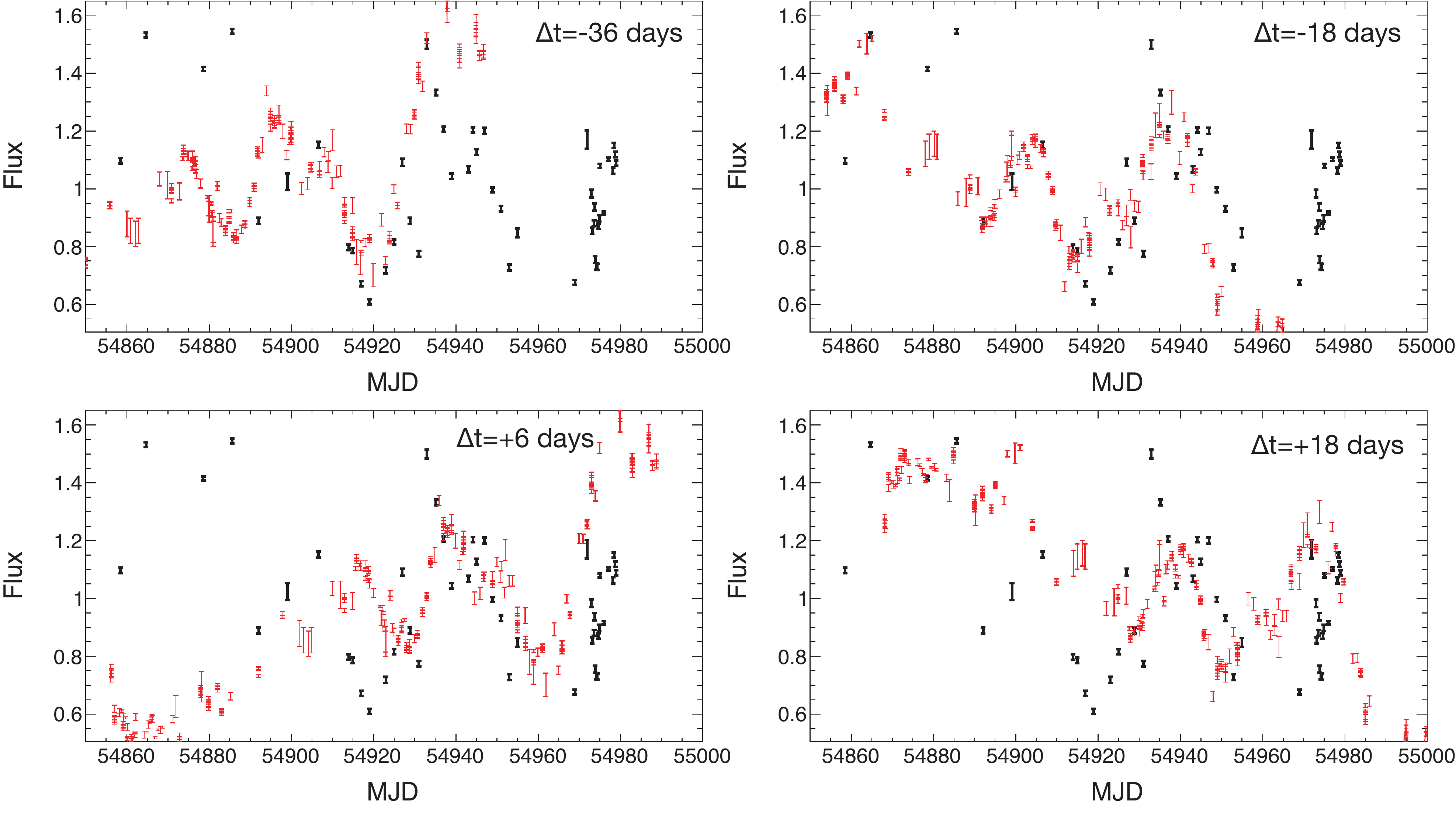}
  \caption{Example plots to illustrate the correlation
    between the optical and the X-ray flux variations. All four panels
    show the normalized \emph{Swift}/XRT (0.3-2~keV) light curve in black. The GASP
    $R$-band light curve, normalized and rescaled to match the same
    flux range as the \emph{Swift}/XRT light curve, is overplotted in
    red with different time lags $\Delta t=-36$, $-18$, $+6$, and $+18$
    days. In case of anti-correlation ($\Delta t=-18$ and $+18$ days),
    the GASP light curve is also flipped vertically.}
\label{fig:dcf_xopt_overplot}
\end{figure*}

Hence we detrended the light curves by fitting and subtracting a
first-order polynomial to each light curve and re-calculated the DCFs
and the ZDCFs. They are shown in panel B of Figure \ref{fig:dcf_xopt} in
comparison to the correlations of the original light curves (panel
A). In the detrended light curves we find two results: 1) The overall
negative correlation spread over most time lags disappears. 2) Some
peaks become evident at time lags $\Delta t\approx-36$, $-18$, $+6$ and
$+18$~days (the latter two being absent in the RXTE/PCA -- UV/optical
DCF). In the ZDCFs these features are generally less pronounced. The
presence of such features leads to the suspicion that an underlying
quasi-periodic behavior may be responsible for the
(anti-)correlations. Indeed there are several local peaks and minima
in both the X-ray and UV/optical light curves. In
Figure \ref{fig:dcf_xopt_overplot} we illustrate how well these features
correlate by overplotting two light curves (\emph{Swift}/XRT (0.3-2~keV) and
GASP $R$ band). Both light curves are normalized. The GASP light curve
is also rescaled such that both light curves cover approximately the
same normalized flux range. In addition, the GASP light curve is
shifted in $x$ direction by $-36$, $-18$, $6$ and $18$ days. For time
lags where an anti-correlation was detected ($-18$ and $+18$~days), we
also flipped the GASP light curve about the horizontal axis (such that
light curve peaks become troughs and vice versa), such that in each
panel of Figure \ref{fig:dcf_xopt_overplot} we should see that both
light curves follow the same path whenever there is a real
(anti-correlation) present. However, it is obvious from these plots
that some, but not all of these features are loosely correlated (as indicated 
by the low statistical significance of the DCF peaks) and
that the limited time window hampers the ability to detect a
convincing correlation. The behavior illustrated in
Figure \ref{fig:dcf_xopt_overplot} may well happen just by chance
without being caused by an underlying physical mechanism. In
Section \ref{sec:periodicities} we show that there is no hint of a
periodic signal in any of the light curves.

\subsection{VHE --- optical/UV}

\begin{figure*}
  \includegraphics[width=17cm]{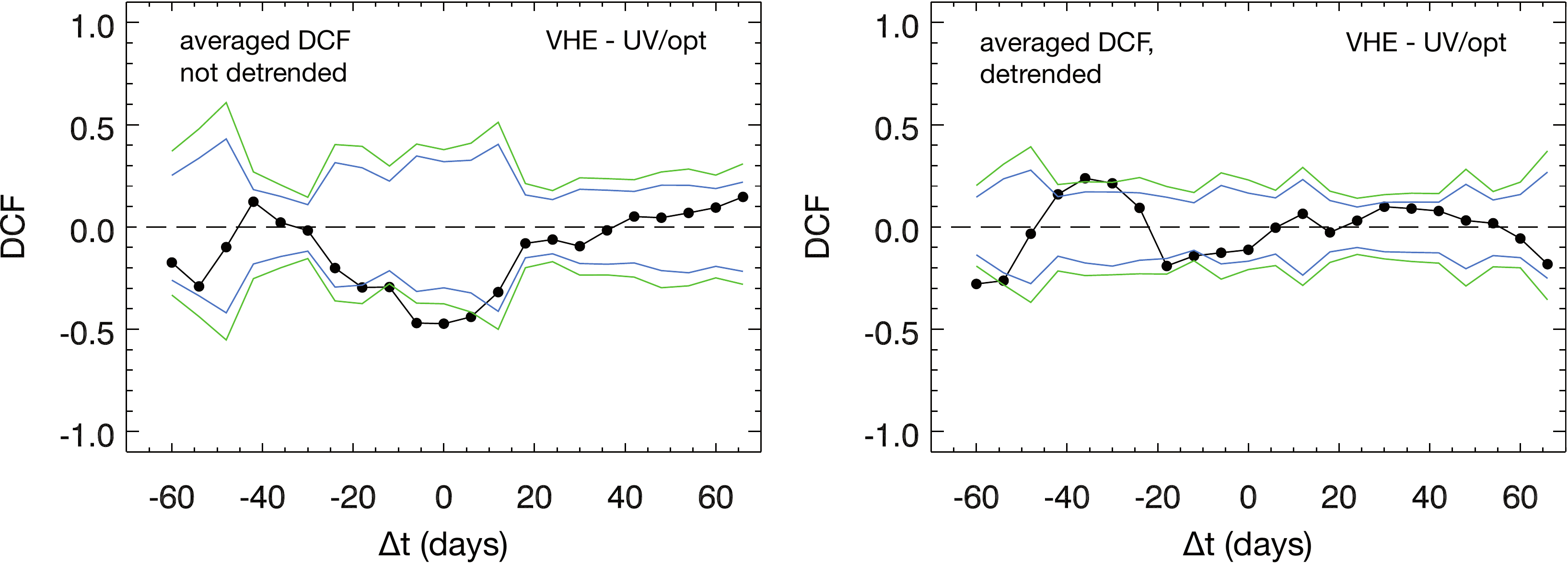}
  \caption{The average of the DCFs of the combined Whipple and MAGIC
    (VHE) light curve, correlated with several optical and UV
    light curves, are shown in black in the left panel. The green
    lines represent the 1\% and 99\% extremes of the likewise averaged
    DCF distribution of simulated optical/UV light curves when
    correlated with the observed VHE light curve. The blue lines
    represent the 5\% and 95\% extremes. The right panel shows the
    same as the left panel, but all light curves have been detrended
    (i.e., fitted and subtracted by a first-order polynomial) before
    correlation.}
\label{fig:dcf_vheopt}
\end{figure*}    

The VHE light curves, when correlated with UV and optical light curves,
produce a strong negative peak at time lag $\Delta t=0$~days
(Figure \ref{fig:dcf_vheopt}), i.e., they are anti-correlated with a
probability larger than $99$\%. However, the optical/UV light curves show a
strong positive trend while the VHE light curves show a weak negative
trend. After detrending the light curves, the anti-correlation at time
lag $\Delta t=0$~days disappears. Instead, the DCF now shows a similar
behavior as the X-ray--optical DCFs (marginally significant
anti-correlation at time-lag $\Delta t\approx-18$~days and correlation at
$\approx-36$~days). This is not surprising given the positive
correlation between VHE and X-ray fluxes. More data in the typical state are
needed in order to judge whether this behavior is just a chance
(anti-)correlation or if it is caused by underlying physical
mechanisms.

\subsection{Other correlations}
All the data in the optical and UV bands vary simultaneously as shown
in Figure \ref{fig:lightcurves}. The NIR bands seem to be well
correlated with the optical and UV bands. However, the number of flux
measurements per NIR light curve is too small to calculate a meaningful
DCF or ZDCF.

No significant correlations are found between radio or HE $\gamma$ rays with
other wavelengths.

\section{Periodicities}\label{sec:periodicities}

\subsection{Lomb-Scargle periodogram}
Although the Lomb-Scargle periodogram (LSP) is not the best way to
determine the PSD for red-noise data \citep{Kastendieck-2011}, it is
a good way to find periodicities when dealing with unevenly sampled
light curves. 

A peak in the LSP at a certain time lag can mean that there is a
periodicity. However, the sampling also produces peaks, e.g., if there
is a flux measurement every $2$ days, there will be a strong peak at
period $P=2$~days. Uneven sampling may result in one or more peaks, if
at least part of the flux measurements follow an approximately regular
observation schedule. The LSPs were determined for periods $\leq
L/5\approx25$~days, where $L$ is the length of the light curve. To
estimate the significance of potential LSP peaks, we also calculated
the LSP for $1000$ simulated light curves each. The simulated light curves
were produced in the same way as in Section \ref{sec:psd} and have the
same underlying PSD (estimated above with PSRESP) and the same
sampling as the original light curve. From the distribution of LSPs we
determined the $95$\% and $99$\% confidence limits. We did not find
significant LSP peaks in any of the light curves. A peak around
$P\approx18$~days is present in a few optical and X-ray LSPs, but
always below $99$\% confidence level, and in most LSPs even below
$95$\% confidence level.

\subsection{Autocorrelation}\label{sec:ACF}

\begin{figure*}
  \includegraphics[width=17cm]{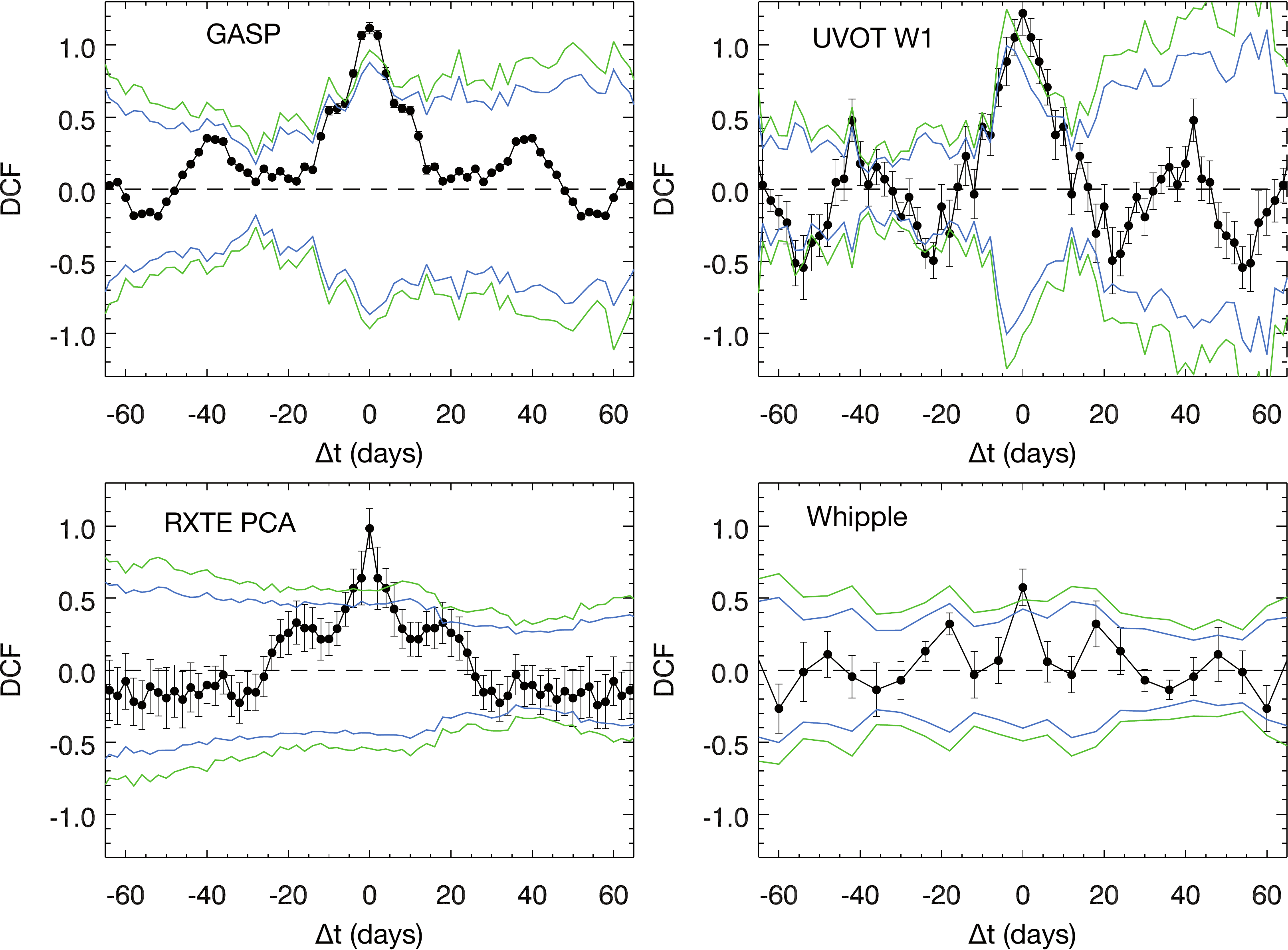}
  \caption{The discrete auto-correlation function for a few
    light curves are shown in black. The green lines represent the
    1\%and 99\% extremes of the likewise averaged DACF distribution of
    simulated light curves when correlated with themselves. The blue
    lines represent the 5\% and 95\% extremes.} 
\label{fig:dacf_periodicities}
\end{figure*}

We use the discrete correlation function \citep{Edelson-1988} to
calculate the discrete auto-correlation function (DACF) of the
variability of Mrk\,421 in all observed wave bands. Equally spaced and
repeated features in the DACF might be a hint to characteristic
timescales and quasi-periodicities. As in the previous sections, we
use simulated light curves to estimate the significance of DACF
peaks, i.e., the observed light curve is correlated with $1000$ simulated 
light curves. This results in confidence limits that are not symmetric around 
zero, although the DACF itself is symmetric. The origin of the asymmetry 
relies on the process used to determine the confidence intervals, which uses 
1000 Monte Carlo realizations of one light curve, together with the asymmetry 
of some of the light curves. This results in different Monte Carlo realizations when 
the light curve is truncated on the left or on the right (negative or positive time lags), 
hence yielding different results for the confidence intervals. Consequently, the asymmetry 
in the confidence intervals is particularly strong where the sampling and variability 
of the light curve changes significantly with time (e.g., UVOT). 
Figure \ref{fig:dacf_periodicities} shows DACFs for a few
representative light curves. There are secondary peaks in some
DACFs. However, they are all well below the 95\% limit, i.e., they do
not appear to be significant. Hence we do not find significant
periodicities or characteristic timescales.

\section{Discussion of the main observational results}\label{sec:results}
Even though Mrk\,421 is known for extreme X-ray and VHE variability,
with short and intense flares
\citep[e.g.][]{Gaidos-1996,Fossati-2008,Aleksic-2012}, the X-ray and
VHE activity measured in the 2009 observing campaign was 
relatively mild, with X-ray/VHE flux variations typically smaller than a
factor of 2. The VHE flux of Mrk\,421 was also relatively low, with
an average flux of about 0.5 times the flux of the Crab
Nebula, which is typical for this source \citep{Acciari-2014}. Regardless 
of the low activity, Mrk\,421 showed significant
variability in the portions of the electromagnetic spectrum where it
emits most of its energy power, namely optical/UV, X-rays and HE/VHE
$\gamma$ rays. The optical/X-ray bands bring information from the
rising/falling segments of the low-energy bump, while the HE/VHE bands
tell us about the rising/falling segment of the high-energy bump. As
reported in Section \ref{sec:Fvar} (see Figure~\ref{fig:Fvar}), the
highest variability occurs at X-rays ($F_{var} \sim$ 0.5), then VHE
($F_{var} \sim$ 0.3), and then optical/UV/HE ($F_{var} \sim$ 0.2). It
is interesting to compare these results with the ones reported
recently for Mrk\,501 \citep{Doert-2013,Mrk501MW2008}, where the fractional
variability increases with energy and is largest at VHE,
instead of X-rays. The comparison of these two observations indicates
that there are fundamental differences in the underlying particle
populations and/or processes producing the broadband radiation in
these two archetypical VHE blazars. Within the one-zone synchrotron
self-Compton scenario, which is commonly used to model the emission of
VHE blazars, the X-ray and VHE variability is driven by the dynamics
of the population of relativistic electrons through their synchrotron
and inverse-Compton emission, respectively. Within this scenario, and
for typical model parameters, the $\sim$keV emission is dominated by
higher-energy electrons, whereas the $\sim$100~GeV emission is
produced by a mixture of lower-energy electrons that inverse-Compton
scatter in Thomson regime, and high-energy electrons that
inverse-Compton scatter in Klein-Nishina regime (see
\citetalias{Abdo-2011a}).
Consequently, the multi-band fractional variability
reported in Section \ref{sec:Fvar} indicates that the population of
higher-energy electrons varies more than 
that at lower energies. 

It is worth noticing that the fractional variabilities detected in the
energy range 2-10 keV measured by \emph{RXTE}/PCA, \emph{RXTE}/ASM and
\emph{Swift}/XRT agree reasonably well with a value of
0.4--0.5\footnote{As discussed in Section \ref{sec:Fvar}, the somewhat
lower value of \emph{RXTE}/ASM, 0.33+/-0.03, is due to the 7-day
integration time, which prevents the detection of variability with
temporal scales of days.} despite the different observing windows of
these three different instruments.  On the other hand, the fractional
variability measured by \emph{Swift}/XRT in the energy range 0.3-2 keV
is $\sim$0.25, which is a factor of $2$ lower than the variability
detected by \emph{Swift}/XRT in the 2-10 keV energy range. Given that
these two observations are performed with the same instrument, the
difference in the fractional variability cannot be ascribed to a
different observing temporal period that might include or exclude a
particular flux variation, and hence the higher variability in the
2-10 keV energy range, in comparison to that in the 0.3-2 keV energy
range, is an intrinsic property of Mrk\,421 during the 2009
observing campaign, which has also been recently reported for Mrk\,421 
during high X-ray and VHE activity observed in 2010 \citep{Aleksic-2015b}. 
Because the characteristic
synchrotron frequency of relativistic electrons is proportional to the
square of the energy of the electrons ($\nu_c \propto E^{2}_{e}$), the
higher synchrotron energies will be produced by higher energy
electrons, and hence this further supports the theoretical framework
of higher variability in the number of higher energy electrons. Given
that the high energy electrons are the ones losing their energy
fastest ($\tau_{cool} \propto E^{-1}_e$), in order to keep the source
emitting X-rays, injection (acceleration) of electrons up to the
highest energies is needed. We therefore conclude that the injection
(acceleration) of high-energy electrons is likely to be the origin of
the flux variations in Mrk\,421.

From the multi-instrument light curves shown in
Figure~\ref{fig:lightcurves}, it can be seen that, while the
variability in X-ray and VHE occurs mostly on $\sim$day timescales,
the variability at optical/UV occurs mostly on $\sim$week or even
longer timescales\footnote{Because of the 3-day span and the relatively 
large statistical uncertainty in the flux points, we cannot evaluate whether
short or long timescales dominate the variability in the HE $\gamma$-ray 
band.}. The different variability timescales do not
show up in the results reported in Section \ref{sec:psd} (see Table
\ref{tab:psdslope}). However, this might be the result of the limited
sensitivity of the PSD analysis due to the uneven sampling of the
light curves, and the rather small range of frequencies sampled
(10$^{-7}$--$\approx10^{-5.7}$ s$^{-1}$), which do not provide a
 long enough lever arm to determine accurately (and ultimately to
distinguish) the index of the power-law spectrum of the PSDs from the
different energy bands. Therefore, the multi-band light curves and
fractional variability show similarities in the X-ray and VHE flux
variations, which differ from characteristics of the optical/UV flux
variations. This observation is further confirmed by the
cross-correlation results reported in Section \ref{sec:correlations},
which show a positive correlation with no time lag between the X-ray
and VHE emission, but not between the optical/UV and X-ray or VHE.
This result indicates that both the X-ray and VHE emissions are
co-spatial and produced by the same population of high-energy
particles. It is worth noting that, while such a correlation has been
reported many times for Mrk\,421 during flaring activity
\citep[e.g.][]{Fossati-2008}, this is the first time that this is
observed during a non-flaring (typical) state. Therefore,
together with the observed harder-when-brighter
behavior in the X-ray spectra, we
interpret this experimental observation as evidence that the
mechanisms responsible for the X-ray/VHE emission during
non-flaring-activity states might not differ substantially from the
ones responsible for the emission during flaring-activity states. In
particular, the positive X-ray/VHE correlation observed during flaring
activity in Mrk\,421 and many other blazars has been interpreted by
many authors as evidence for leptonic scenarios, and hence against
the hadronic scenarios where the X-ray emission and the $\gamma$-ray
emission are produced by different particle populations and
processes. However, with a fine tuning of the parameters, hadronic models
are also able to explain single flaring events with an X-ray/VHE
correlation with time lag zero \citep{Mastichiadis-2013}.

The observations presented here confirm that the relation between
X-ray and VHE bands also exists when the source is not flaring. That
is, such a relation is persistent over long timescales of at least
several months, and does not occur only on single flaring events, and
that is much more difficult to explain with hadronic
scenarios. Therefore, these observations provide strong 
evidence supporting leptonic scenarios as responsible for the dominant
X-ray/VHE emission from Mrk\,421. 

Another result that is worth discussing is the overall
anti-correlation between the X-ray and optical bands. This
anti-correlation spreads over a large range of time lags and hence it
is fundamentally different from the one obtained for the X-ray and VHE
bands. As we show in Section \ref{sec:correlations}, the origin of
this overall anti-correlation is the long-term trends of the
optical/UV and X-ray activity: while the former increases over time
during the observing campaign, the latter decreases. 
The temporal evolution of the optical
and X-ray/VHE bands is complex, and a dedicated correlation analysis
over many years will be necessary in order to properly characterize
it.

For the 2009 multi-wavelength campaign, we do observe an
anti-correlation in the (long-term) temporal evolution between
optical/UV and X-rays, and hence it is worth discussing possible
theoretical scenarios that might lead to this situation. The first
scenario is that the optical/UV and the X-ray/VHE emissions are
dominated by the emission from two distinct and unconnected regions
with different temporal evolutions of their respective particle
populations. In such case, the optical/UV vs. X-ray anti-correlation
observed in the 2009 multi-instrument campaign would have occurred by
chance, and hence we would also expect to see multi-month time
intervals with a positive correlation, or no correlation. A second
scenario would be a two-component (high- and low-energy) particle
population, in which the low- and high-energy particles have a
different but related long-term temporal evolution. A change in the
magnetic field intensity while keeping the acceleration timescale
constant could lead to the observed optical/UV--X-ray (long-term)
anti-correlation. An increase in the magnetization would produce a
higher synchrotron emission with a decrease in the energy of the 
electrons (due to a stronger cooling), which effectively
would lead to a higher optical emission with a lower X-ray
emission. On the other hand, a lower magnetization would lead to a
decrease in the total emitted synchrotron flux, but a higher maximum
electron energy, which effectively could lower the optical flux and
increase the X-ray flux. In practice, such a scenario would be
somewhat similar to the blazar sequence \citep{Ghisellini-1998}, with
the difference that in the latter scenario the different coolings
would relate to different sources instead of different states of the
same source. A third scenario could be a global long-term change in
the efficiency of the acceleration mechanism that produces the
electron energy distribution. Such a change in the global efficiency
could shift the entire synchrotron bump to higher/lower energies. For
instance, if the acceleration mechanism becomes more efficient to get
electrons up to the highest energies at the expense of keeping a lower
number of low-energy electrons (i.e., the index of the electron
population gets harder), the emission at the rising segment of the
synchrotron bump (optical) would decrease, while that on the
decreasing segment of the synchrotron bump (X-rays) would increase.

In this study we did not see any correlation between the radio fluxes 
and those at higher frequencies. However, since the measured radio 
emission is expected to have large contributions from regions farther 
away in the jet, it is not surprising to see a non-correlation between 
radio and optical, X-ray and/or $\gamma$-ray energies on timescales of days 
to weeks. We note, however, that such correlation might be apparent 
during large flares when the radio emission might be strongly 
dominated by the same region responsible for the overall broadband 
emission.

\section{Conclusions}

We studied the broadband evolution of the SED of Mrk\,421 through a
4.5~month long multi-instrument observing campaign in 2009, when the
source was in its typical (non-flaring) activity state, with a VHE
flux of about half that of the Crab Nebula. Even though the source
did not show flaring activity, we could measure significant
variability in the energy bands where the emitted power is largest:
optical, X-ray, $\gamma$ rays and VHE. The highest variability occurred in the
X-ray band, which, within the standard one-zone SSC scenario,
indicates that the high-energy electrons are more variable than the
low-energy electrons. We also observed a harder-when-brighter
 behavior in the X-ray spectra, and found a positive correlation between
the X-ray and VHE bands. In the literature one can find many works
reporting a positive correlation between the X-ray and VHE fluxes
\citep[e.g.][and references therein]{Fossati-2008} and spectral
shape changes with the X-ray flux \citep[e.g.][]{Tramacere-2009}, but only when
Mrk\,421 was showing VHE flaring activity (i.e., VHE flux above the
flux of the Crab Nebula). This is the first time that such characteristics
are reported for non-flaring activity and
suggests that the processess occurring during the flaring activity
also occur when the source is in a non-flaring (low) state. In
particular, this is a strong argument in favor of leptonic scenarios
dominating the broadband emission of Mrk\,421 during non-flaring
activity, since such a temporally extended X-ray/VHE correlation
cannot be explained within the standard hadronic scenarios. Moreover,
a negative correlation in the (long-term) temporal evolution of the
optical/UV and X-ray bands was also observed. Such a trend could be
produced in a region with a particle population where the low- and
high-energy particles evolve differently but in a related way, which
could be produced by a change in the magnetization of the region while
keeping the acceleration timescales constant, or by a global change in
the efficiency of the mechanism accelerating the electrons. In any
case, even though statistically significant for the 2009
multi-instrument campaign, the current dataset does not allow us to
exclude that this optical/X-ray anti-correlation
was observed by chance, and hence that the optical and the X-ray bands
are produced by distinct and unrelated particle populations that
evolve separately. Further multi-instrument observations extending
over many years will help to address this question.

\begin{acknowledgements} 

We would like to thank the referee for the useful comments that helped to improve the manuscript. 
We also thank Patricia Ar\'{e}valo for helpful contributions and suggestions.

The MAGIC collaboration would like to thank the Instituto de Astrof\'{\i}sica de Canarias for the excellent working conditions at the Observatorio del Roque de los Muchachos in La Palma. The financial support of the German BMBF and MPG, the Italian INFN and INAF,  the Swiss National Fund SNF, the ERDF under the Spanish MINECO, and the Japanese JSPS and MEXT is gratefully acknowledged. This work was also supported by the Centro de Excelencia Severo Ochoa SEV-2012-0234, CPAN CSD2007-00042 and MultiDark CSD2009-00064 projects of the Spanish Consolider-Ingenio 2010 programme, by grant 268740 of the Academy of Finland, by the Croatian Science Foundation (HrZZ) Project 09/176 and the University of Rijeka Project 13.12.1.3.02, by the DFG Collaborative Research Centers SFB823/C4 and SFB876/C3, and by the Polish MNiSzW grant 745/N-HESS-MAGIC/2010/0.

The VERITAS collaboration acknowledges support from the
U.S. Department of Energy, the U.S. National Science Foundation and
the Smithsonian Institution, by NSERC in Canada, by Science Foundation
Ireland, and by STCF in the UK. We acknowledge the excellent work of
the technical support at the FLWO and the collaboration institutions
in the construction and operation of the instrument.  

The \emph{Fermi}-LAT Collaboration acknowledges support from a number
of agencies and institutes for both development and the operation of
the LAT as well as scientific data analysis. These include NASA and
DOE in the United States, CEA/Irfu and IN2P3/CNRS in France, ASI and
INFN in Italy, MEXT, KEK, and JAXA in Japan, and the K. A. Wallenberg
Foundation, the Swedish Research Council and the National Space Board
in Sweden. Additional support from INAF in Italy and CNES in France
for science analysis during the operations phase is also gratefully
acknowledged.

We acknowledge the use of public data from the \emph{Swift} and RXTE data archives.
The OVRO 40 m monitoring program is
supported in part by NASA grants NNX08AW31G 
and NNX11A043G, and NSF grants AST-0808050 
and AST-1109911.
The Mets\"ahovi team acknowledges the support from the Academy of Finland
to our observing projects (numbers 212656, 210338, 121148, and others).
The Abastumani Observatory team acknowledges financial  support by the Shota
Rustaveli National Science Foundation through project FR/577/6-320/13. The St. Petersburg 
University team acknowledges support from the Russian RFBR foundation via grant 09-02-00092.
AZT-24 observations are made within an agreement between Pulkovo, Rome and Teramo 
observatories. This research is partly based on observations with the 100 m telescope of the MPIfR
(Max-Planck-Institut fuer Radioastronomie) at Effelsberg, as well as with the Medicina and Noto 
telescopes operated by INAF--Istituto di Radioastronomia. 
M. Villata organized the optical-to-radio observations by GASP-WEBT as the president of the
collaboration.

\end{acknowledgements}

\bibliographystyle{aa}
\bibliography{aa24216_arxiv}

\begin{thebibliography}{45}
\expandafter\ifx\csname natexlab\endcsname\relax\def\natexlab#1{#1}\fi

\bibitem[{{Abdo} {et~al.}(2010){Abdo}, {Ackermann}, {Ajello}, {Antolini},
  {Baldini}, {Ballet}, {Barbiellini}, {Bastieri}, {Bechtol}, {Bellazzini},
  {Berenji}, {Blandford}, {Bloom}, {Bonamente}, {Borgland}, {Bouvier},
  {Bregeon}, {Brez}, {Brigida}, {Bruel}, {Buehler}, {Burnett}, {Buson},
  {Caliandro}, {Cameron}, {Caraveo}, {Carrigan}, {Casandjian}, {Cavazzuti},
  {Cecchi}, {{\c C}elik}, {Chekhtman}, {Cheung}, {Chiang}, {Ciprini}, {Claus},
  {Cohen-Tanugi}, {Cominsky}, {Conrad}, {Costamante}, {Cutini}, {Dermer}, {de
  Angelis}, {de Palma}, {Silva}, {Drell}, {Dubois}, {Dumora}, {Farnier},
  {Favuzzi}, {Fegan}, {Focke}, {Fortin}, {Frailis}, {Fukazawa}, {Funk},
  {Fusco}, {Gargano}, {Gasparrini}, {Gehrels}, {Germani}, {Giebels},
  {Giglietto}, {Giommi}, {Giordano}, {Glanzman}, {Godfrey}, {Grenier},
  {Grondin}, {Grove}, {Guiriec}, {Hadasch}, {Hayashida}, {Hays}, {Healey},
  {Horan}, {Hughes}, {Itoh}, {J{\'o}hannesson}, {Johnson}, {Johnson}, {Kamae},
  {Katagiri}, {Kataoka}, {Kawai}, {Kn{\"o}dlseder}, {Kuss}, {Lande}, {Larsson},
  {Latronico}, {Lemoine-Goumard}, {Longo}, {Loparco}, {Lott}, {Lovellette},
  {Lubrano}, {Madejski}, {Makeev}, {Massaro}, {Mazziotta}, {McEnery},
  {Michelson}, {Mitthumsiri}, {Mizuno}, {Moiseev}, {Monte}, {Monzani},
  {Morselli}, {Moskalenko}, {Mueller}, {Murgia}, {Nolan}, {Norris}, {Nuss},
  {Ohno}, {Ohsugi}, {Omodei}, {Orlando}, {Ormes}, {Ozaki}, {Panetta}, {Parent},
  {Pelassa}, {Pepe}, {Pesce-Rollins}, {Piron}, {Porter}, {Rain{\`o}}, {Rando},
  {Razzano}, {Reimer}, {Reimer}, {Ritz}, {Rodriguez}, {Romani}, {Roth}, {Ryde},
  {Sadrozinski}, {Sander}, {Scargle}, {Sgr{\`o}}, {Shaw}, {Smith}, {Spandre},
  {Spinelli}, {Starck}, {Strickman}, {Suson}, {Takahashi}, {Takahashi},
  {Tanaka}, {Thayer}, {Thayer}, {Thompson}, {Tibaldo}, {Torres}, {Tosti},
  {Tramacere}, {Uchiyama}, {Usher}, {Vasileiou}, {Vilchez}, {Vitale}, {Waite},
  {Wallace}, {Wang}, {Winer}, {Wood}, {Yang}, {Ylinen}, \&
  {Ziegler}}]{Abdo-2010}
{Abdo}, A.~A., {Ackermann}, M., {Ajello}, M., {et~al.} 2010, \apj, 722, 520

\bibitem[{{Abdo} {et~al.}(2011{\natexlab{a}}){Abdo}, {Ackermann}, {Ajello},
  {Allafort}, {Baldini}, {Ballet}, {Barbiellini}, {Baring}, {Bastieri},
  {Bechtol}, \& et~al.}]{Abdo-2011b}
{Abdo}, A.~A., {Ackermann}, M., {Ajello}, M., {et~al.} 2011{\natexlab{a}},
  \apj, 727, 129

\bibitem[{{Abdo} {et~al.}(2011{\natexlab{b}}){Abdo}, {Ackermann}, {Ajello},
  {Baldini}, {Ballet}, {Barbiellini}, {Bastieri}, {Bechtol}, {Bellazzini},
  {Berenji}, \& et~al.}]{Abdo-2011a}
{Abdo}, A.~A., {Ackermann}, M., {Ajello}, M., {et~al.} 2011{\natexlab{b}},
  \apj, 736, 131

\bibitem[{{Acciari} {et~al.}(2014){Acciari}, {Arlen}, {Aune}, {Benbow}, {Bird},
  {Bouvier}, {Bradbury}, {Buckley}, {Bugaev}, {de la Calle Perez},
  {Carter-Lewis}, {Cesarini}, {Ciupik}, {Collins-Hughes}, {Connolly}, {Cui},
  {Duke}, {Dumm}, {Falcone}, {Federici}, {Fegan}, {Fegan}, {Finley},
  {Finnegan}, {Fortson}, {Gaidos}, {Galante}, {Gall}, {Gibbs}, {Gillanders},
  {Griffin}, {Grube}, {Gyuk}, {Hanna}, {Horan}, {Humensky}, {Kaaret},
  {Kertzman}, {Khassen}, {Kieda}, {Krawczynski}, {Krennrich}, {Lang},
  {McEnery}, {Madhavan}, {Moriarty}, {Nelson}, {O'Faol{\'a}in de Bhr{\'o}ithe},
  {Ong}, {Orr}, {Otte}, {Perkins}, {Petry}, {Pichel}, {Pohl}, {Quinn}, {Ragan},
  {Reynolds}, {Roache}, {Rovero}, {Schroedter}, {Sembroski}, {Smith},
  {Telezhinsky}, {Theiling}, {Toner}, {Tyler}, {Varlotta}, {Vivier}, {Wakely},
  {Ward}, {Weekes}, {Weinstein}, {Welsing}, {Williams}, \&
  {Wissel}}]{Acciari-2014}
{Acciari}, V.~A., {Arlen}, T., {Aune}, T., {et~al.} 2014, Astroparticle
  Physics, 54, 1

\bibitem[{{Albert} {et~al.}(2007){Albert}, {Aliu}, {Anderhub}, {Antoranz},
  {Armada}, {Asensio}, {Baixeras}, {Barrio}, {Bartko}, {Bastieri}, {Becker},
  {Bednarek}, {Berger}, {Bigongiari}, {Biland}, {Bock}, {Bordas},
  {Bosch-Ramon}, {Bretz}, {Britvitch}, {Camara}, {Carmona}, {Chilingarian},
  {Ciprini}, {Coarasa}, {Commichau}, {Contreras}, {Cortina}, {Curtef},
  {Danielyan}, {Dazzi}, {De Angelis}, {de los Reyes}, {De Lotto},
  {Domingo-Santamar{\'{\i}}a}, {Dorner}, {Doro}, {Errando}, {Fagiolini},
  {Ferenc}, {Fern{\'a}ndez}, {Firpo}, {Flix}, {Fonseca}, {Font}, {Fuchs},
  {Galante}, {Garczarczyk}, {Gaug}, {Giller}, {Goebel}, {Hakobyan},
  {Hayashida}, {Hengstebeck}, {H{\"o}hne}, {Hose}, {Hsu}, {Jacon}, {Jogler},
  {Kalekin}, {Kosyra}, {Kranich}, {Kritzer}, {Laatiaoui}, {Laille}, {Liebing},
  {Lindfors}, {Lombardi}, {Longo}, {L{\'o}pez}, {L{\'o}pez}, {Lorenz},
  {Majumdar}, {Maneva}, {Mannheim}, {Mansutti}, {Mariotti}, {Mart{\'{\i}}nez},
  {Mazin}, {Merck}, {Meucci}, {Meyer}, {Miranda}, {Mirzoyan}, {Mizobuchi},
  {Moralejo}, {Nilsson}, {Ninkovic}, {O{\~n}a-Wilhelmi}, {Ordu{\~n}a}, {Otte},
  {Oya}, {Paneque}, {Paoletti}, {Paredes}, {Pasanen}, {Pascoli}, {Pauss},
  {Pegna}, {Persic}, {Peruzzo}, {Piccioli}, {Poller}, {Prandini}, {Raymers},
  {Rhode}, {Rib{\'o}}, {Rico}, {Rissi}, {Robert}, {R{\"u}gamer}, {Saggion},
  {S{\'a}nchez}, {Sartori}, {Scalzotto}, {Scapin}, {Schmitt}, {Schweizer},
  {Shayduk}, {Shinozaki}, {Shore}, {Sidro}, {Sillanp{\"a}{\"a}}, {Sobczynska},
  {Stamerra}, {Stark}, {Takalo}, {Temnikov}, {Tescaro}, {Teshima}, {Tonello},
  {Torres}, {Torres}, {Turini}, {Vankov}, {Vitale}, {Wagner}, {Wibig},
  {Wittek}, {Zanin}, \& {Zapatero}}]{Albert-2007}
{Albert}, J., {Aliu}, E., {Anderhub}, H., {et~al.} 2007, \apj, 663, 125

\bibitem[{{Aleksi{\'c}} {et~al.}(2012){Aleksi{\'c}}, {Alvarez}, {Antonelli},
  {Antoranz}, {Asensio}, {Backes}, {Barrio}, {Bastieri}, {Becerra
  Gonz{\'a}lez}, {Bednarek}, {Berdyugin}, {Berger}, {Bernardini}, {Biland},
  {Blanch}, {Bock}, {Boller}, {Bonnoli}, {Borla Tridon}, {Braun}, {Bretz},
  {Ca{\~n}ellas}, {Carmona}, {Carosi}, {Colin}, {Colombo}, {Contreras},
  {Cortina}, {Cossio}, {Covino}, {Dazzi}, {De Angelis}, {De Caneva}, {De Cea
  del Pozo}, {De Lotto}, {Delgado Mendez}, {Diago Ortega}, {Doert},
  {Dom{\'{\i}}nguez}, {Dominis Prester}, {Dorner}, {Doro}, {Elsaesser},
  {Ferenc}, {Fonseca}, {Font}, {Fruck}, {Garc{\'{\i}}a L{\'o}pez},
  {Garczarczyk}, {Garrido}, {Giavitto}, {Godinovi{\'c}}, {Hadasch},
  {H{\"a}fner}, {Herrero}, {Hildebrand}, {H{\"o}hne-M{\"o}nch}, {Hose},
  {Hrupec}, {Huber}, {Jogler}, {Kellermann}, {Klepser}, {Kr{\"a}henb{\"u}hl},
  {Krause}, {La Barbera}, {Lelas}, {Leonardo}, {Lindfors}, {Lombardi},
  {L{\'o}pez}, {L{\'o}pez}, {Lorenz}, {Makariev}, {Maneva}, {Mankuzhiyil},
  {Mannheim}, {Maraschi}, {Mariotti}, {Mart{\'{\i}}nez}, {Mazin}, {Meucci},
  {Miranda}, {Mirzoyan}, {Miyamoto}, {Mold{\'o}n}, {Moralejo}, {Munar-Adrover},
  {Nieto}, {Nilsson}, {Orito}, {Oya}, {Paneque}, {Paoletti}, {Pardo},
  {Paredes}, {Partini}, {Pasanen}, {Pauss}, {Perez-Torres}, {Persic},
  {Peruzzo}, {Pilia}, {Pochon}, {Prada}, {Prada Moroni}, {Prandini}, {Puljak},
  {Reichardt}, {Reinthal}, {Rhode}, {Rib{\'o}}, {Rico}, {R{\"u}gamer},
  {Saggion}, {Saito}, {Saito}, {Salvati}, {Satalecka}, {Scalzotto}, {Scapin},
  {Schultz}, {Schweizer}, {Shayduk}, {Shore}, {Sillanp{\"a}{\"a}}, {Sitarek},
  {Sobczynska}, {Spanier}, {Spiro}, {Stamerra}, {Steinke}, {Storz}, {Strah},
  {Suri{\'c}}, {Takalo}, {Takami}, {Tavecchio}, {Temnikov}, {Terzi{\'c}},
  {Tescaro}, {Teshima}, {Tibolla}, {Torres}, {Treves}, {Uellenbeck}, {Vankov},
  {Vogler}, {Wagner}, {Weitzel}, {Zabalza}, {Zandanel}, \&
  {Zanin}}]{Aleksic-2012}
{Aleksi{\'c}}, J., {Alvarez}, E.~A., {Antonelli}, L.~A., {et~al.} 2012, \aap,
  542, A100

\bibitem[{{Aleksi{\'c}} {et~al.}(2015{\natexlab{a}}){Aleksi{\'c}}, {Ansoldi}, {Antonelli},
  {Antoranz}, {Babic}, {Bangale}, {Barres de Almeida}, {Barrio}, {Becerra
  Gonz{\'a}lez}, {Bednarek}, {Berger}, {Bernardini}, {Biland}, {Blanch},
  {Bock}, {Bonnefoy}, {Bonnoli}, {Borracci}, {Bretz}, {Carmona}, {Carosi},
  {Carreto Fidalgo}, {Colin}, {Colombo}, {Contreras}, {Cortina}, {Covino}, {da
  Vela}, {Dazzi}, {de Angelis}, {de Caneva}, {de Lotto}, {Delgado Mendez},
  {Doert}, {Dom{\'{\i}}nguez}, {Dominis Prester}, {Dorner}, {Doro}, {Einecke},
  {Eisenacher}, {Elsaesser}, {Farina}, {Ferenc}, {Fonseca}, {Font}, {Frantzen},
  {Fruck}, {Garc{\'{\i}}a L{\'o}pez}, {Garczarczyk}, {Garrido Terrats}, {Gaug},
  {Giavitto}, {Godinovi{\'c}}, {Gonz{\'a}lez Mu{\~n}oz}, {Gozzini}, {Hadamek},
  {Hadasch}, {Herrero}, {Hildebrand}, {Hose}, {Hrupec}, {Idec}, {Kadenius},
  {Kellermann}, {Knoetig}, {Krause}, {Kushida}, {La Barbera}, {Lelas},
  {Lewandowska}, {Lindfors}, {Lombardi}, {L{\'o}pez}, {L{\'o}pez-Coto},
  {L{\'o}pez-Oramas}, {Lorenz}, {Lozano}, {Makariev}, {Mallot}, {Maneva},
  {Mankuzhiyil}, {Mannheim}, {Maraschi}, {Marcote}, {Mariotti},
  {Mart{\'{\i}}nez}, {Mazin}, {Menzel}, {Meucci}, {Miranda}, {Mirzoyan},
  {Moralejo}, {Munar-Adrover}, {Nakajima}, {Niedzwiecki}, {Nilsson}, {Nowak},
  {Orito}, {Overkemping}, {Paiano}, {Palatiello}, {Paneque}, {Paoletti},
  {Paredes}, {Paredes-Fortuny}, {Partini}, {Persic}, {Prada}, {Prada Moroni},
  {Prandini}, {Preziuso}, {Puljak}, {Reinthal}, {Rhode}, {Rib{\'o}}, {Rico},
  {Rodriguez Garcia}, {R{\"u}gamer}, {Saggion}, {Saito}, {Saito}, {Salvati},
  {Satalecka}, {Scalzotto}, {Scapin}, {Schultz}, {Schweizer}, {Shore},
  {Sillanp{\"a}{\"a}}, {Sitarek}, {Snidaric}, {Sobczynska}, {Spanier},
  {Stamatescu}, {Stamerra}, {Steinbring}, {Storz}, {Sun}, {Suri{\'c}},
  {Takalo}, {Tavecchio}, {Temnikov}, {Terzi{\'c}}, {Tescaro}, {Teshima},
  {Thaele}, {Tibolla}, {Torres}, {Toyama}, {Treves}, {Uellenbeck}, {Vogler},
  {Wagner}, {Zandanel}, {Zanin}, \& {MAGIC Collaboration}}]{Mrk501MW2008}
{Aleksi{\'c}}, J., {Ansoldi}, S., {Antonelli}, L.~A., {et~al.} 2015{\natexlab{a}}, \aap, 573,
  A50

\bibitem[{{Aleksi{\'c}} {et~al.}(2015{\natexlab{b}}){Aleksi{\'c}}, {Ansoldi}, {Antonelli},
  {Antoranz}, {Babic}, {Bangale}, {Barres de Almeida}, {Barrio}, {Becerra
  Gonz{\'a}lez}, {et~al.}}]{Aleksic-2015b}
{Aleksi{\'c}}, J., {Ansoldi}, S., {Antonelli}, L.~A., {et~al.} 2015{\natexlab{b}},
  arXiv:astro-ph/1412.3576

\bibitem[{{Alexander}(1997)}]{Alexander-1997}
{Alexander}, T. 1997, in Astrophysics and Space Science Library, Vol. 218,
  Astronomical Time Series, ed. D.~{Maoz}, A.~{Sternberg}, \& E.~M.
  {Leibowitz}, 163

\bibitem[{{Ar{\'e}valo} {et~al.}(2009){Ar{\'e}valo}, {Uttley}, {Lira},
  {Breedt}, {McHardy}, \& {Churazov}}]{Arevalo-2009}
{Ar{\'e}valo}, P., {Uttley}, P., {Lira}, P., {et~al.} 2009, \mnras, 397, 2004

\bibitem[{{Balokovic} {et~al.}(2013){Balokovic}, {Furniss}, {Madejski}, \&
  {Harrison}}]{Balokovic-2013}
{Balokovic}, M., {Furniss}, A., {Madejski}, G., \& {Harrison}, F. 2013, The
  Astronomer's Telegram, 4974, 1

\bibitem[{{Chatterjee} {et~al.}(2012){Chatterjee}, {Bailyn}, {Bonning},
  {Buxton}, {Coppi}, {Fossati}, {Isler}, {Maraschi}, \&
  {Urry}}]{Chatterjee-2012}
{Chatterjee}, R., {Bailyn}, C.~D., {Bonning}, E.~W., {et~al.} 2012, \apj, 749,
  191

\bibitem[{{Chatterjee} {et~al.}(2008){Chatterjee}, {Jorstad}, {Marscher}, {Oh},
  {McHardy}, {Aller}, {Aller}, {Balonek}, {Miller}, {Ryle}, {Tosti},
  {Kurtanidze}, {Nikolashvili}, {Larionov}, \& {Hagen-Thorn}}]{Chatterjee-2008}
{Chatterjee}, R., {Jorstad}, S.~G., {Marscher}, A.~P., {et~al.} 2008, \apj,
  689, 79

\bibitem[{{Cortina} \& {Holder}(2013)}]{Cortina-2013}
{Cortina}, J. \& {Holder}, J. 2013, The Astronomer's Telegram, 4976, 1

\bibitem[{{Dietrich} {et~al.}(1998){Dietrich}, {Peterson}, {Albrecht},
  {Altmann}, {Barth}, {Bennie}, {Bertram}, {Bochkarev}, {Bock}, {Braun},
  {Burenkov}, {Collier}, {Fang}, {Francis}, {Filippenko}, {Foltz}, {Gaessler},
  {Gaskell}, {Geffert}, {Ghosh}, {Hilditch}, {Honeycutt}, {Horne}, {Huchra},
  {Kaspi}, {Kuemmel}, {Leighly}, {Leonard}, {Malkov}, {Mikhailov}, {Miller},
  {Morrill}, {Noble}, {O'Brien}, {Oswalt}, {Pebley}, {Pfeiffer}, {Pronik},
  {Qian}, {Robertson}, {Robinson}, {Rumstay}, {Schmoll}, {Sergeev}, {Sergeeva},
  {Shapovalova}, {Skillman}, {Snedden}, {Soundararajaperumal}, {Stirpe}, {Tao},
  {Turner}, {Wagner}, {Wagner}, {Wei}, {Wu}, {Zheng}, \& {Zou}}]{Dietrich-1998}
{Dietrich}, M., {Peterson}, B.~M., {Albrecht}, P., {et~al.} 1998, \apjs, 115,
  185

\bibitem[{{Doert} {et~al.}(2013){Doert}, {David Paneque for the MAGIC
  Collaboration}, {the VERITAS Collaboration}, \& {the Fermi-LAT
  Collaboration}}]{Doert-2013}
{Doert}, M., {David Paneque for the MAGIC Collaboration}, {the VERITAS
  Collaboration}, \& {the Fermi-LAT Collaboration}. 2013,
  arXiv:astro-ph/1307.8344

\bibitem[{{Edelson} \& {Krolik}(1988)}]{Edelson-1988}
{Edelson}, R.~A. \& {Krolik}, J.~H. 1988, \apj, 333, 646

\bibitem[{{Fortson} {et~al.}(2012){Fortson}, {VERITAS Collaboration}, \&
  {Fermi-LAT Collaborators}}]{Fortson-2012}
{Fortson}, L., {VERITAS Collaboration}, \& {Fermi-LAT Collaborators}. 2012, in
  American Institute of Physics Conference Series, Vol. 1505, American
  Institute of Physics Conference Series, ed. F.~A. {Aharonian}, W.~{Hofmann},
  \& F.~M. {Rieger}, 514--517

\bibitem[{{Fossati} {et~al.}(2008){Fossati}, {Buckley}, {Bond}, {Bradbury},
  {Carter-Lewis}, {Chow}, {Cui}, {Falcone}, {Finley}, {Gaidos}, {Grube},
  {Holder}, {Horan}, {Horns}, {Jordan}, {Kieda}, {Kildea}, {Krawczynski},
  {Krennrich}, {Lang}, {LeBohec}, {Lee}, {Moriarty}, {Ong}, {Petry}, {Quinn},
  {Sembroski}, {Wakely}, \& {Weekes}}]{Fossati-2008}
{Fossati}, G., {Buckley}, J.~H., {Bond}, I.~H., {et~al.} 2008, \apj, 677, 906

\bibitem[{{Fossati} {et~al.}(2000){Fossati}, {Celotti}, {Chiaberge}, {Zhang},
  {Chiappetti}, {Ghisellini}, {Maraschi}, {Tavecchio}, {Pian}, \&
  {Treves}}]{Fossati-2000}
{Fossati}, G., {Celotti}, A., {Chiaberge}, M., {et~al.} 2000, \apj, 541, 166

\bibitem[{{Gaidos} {et~al.}(1996){Gaidos}, {Akerlof}, {Biller}, {Boyle},
  {Breslin}, {Buckley}, {Carter-Lewis}, {Catanese}, {Cawley}, {Fegan},
  {Finley}, {Gordo}, {Hillas}, {Krennrich}, {Lamb}, {Lessard}, {McEnery},
  {Masterson}, {Mohanty}, {Moriarty}, {Quinn}, {Rodgers}, {Rose}, {Samuelson},
  {Schubnell}, {Sembroski}, {Srinivasan}, {Weekes}, {Wilson}, \&
  {Zweerink}}]{Gaidos-1996}
{Gaidos}, J.~A., {Akerlof}, C.~W., {Biller}, S., {et~al.} 1996, \nat, 383, 319

\bibitem[{{Ghisellini} {et~al.}(1998){Ghisellini}, {Celotti}, {Fossati},
  {Maraschi}, \& {Comastri}}]{Ghisellini-1998}
{Ghisellini}, G., {Celotti}, A., {Fossati}, G., {Maraschi}, L., \& {Comastri},
  A. 1998, \mnras, 301, 451

\bibitem[{{Graff} {et~al.}(2008){Graff}, {Georganopoulos}, {Perlman}, \&
  {Kazanas}}]{Graff-2008}
{Graff}, P.~B., {Georganopoulos}, M., {Perlman}, E.~S., \& {Kazanas}, D. 2008,
  \apj, 689, 68

\bibitem[{{Hillas} {et~al.}(1998){Hillas}, {Akerlof}, {Biller}, {Buckley},
  {Carter-Lewis}, {Catanese}, {Cawley}, {Fegan}, {Finley}, {Gaidos},
  {Krennrich}, {Lamb}, {Lang}, {Mohanty}, {Punch}, {Reynolds}, {Rodgers},
  {Rose}, {Rovero}, {Schubnell}, {Sembroski}, {Vacanti}, {Weekes}, {West}, \&
  {Zweerink}}]{Hillas-1998}
{Hillas}, A.~M., {Akerlof}, C.~W., {Biller}, S.~D., {et~al.} 1998, \apj, 503,
  744

\bibitem[{{Horan} {et~al.}(2007){Horan}, {Atkins}, {Badran}, {Blaylock},
  {Bradbury}, {Buckley}, {Byrum}, {Celik}, {Chow}, {Cogan}, {Cui}, {Daniel},
  {de la Calle Perez}, {Dowdall}, {Falcone}, {Fegan}, {Fegan}, {Finley},
  {Fortin}, {Fortson}, {Gillanders}, {Grube}, {Gutierrez}, {Hall}, {Hanna},
  {Holder}, {Hughes}, {Humensky}, {Kenny}, {Kertzman}, {Kieda}, {Kildea},
  {Krawczynski}, {Krennrich}, {Lang}, {LeBohec}, {Maier}, {Moriarty}, {Nagai},
  {Ong}, {Perkins}, {Petry}, {Quinn}, {Quinn}, {Ragan}, {Reynolds}, {Rose},
  {Schroedter}, {Sembroski}, {Steele}, {Swordy}, {Toner}, {Valcarcel},
  {Vassiliev}, {Wagner}, {Wakely}, {Weekes}, {White}, \&
  {Williams}}]{Horan-2007}
{Horan}, D., {Atkins}, R.~W., {Badran}, H.~M., {et~al.} 2007, \apj, 655, 396

\bibitem[{{Kastendieck} {et~al.}(2011){Kastendieck}, {Ashley}, \&
  {Horns}}]{Kastendieck-2011}
{Kastendieck}, M.~A., {Ashley}, M.~C.~B., \& {Horns}, D. 2011, \aap, 531, A123

\bibitem[{{Kataoka} {et~al.}(2001){Kataoka}, {Takahashi}, {Wagner}, {Iyomoto},
  {Edwards}, {Hayashida}, {Inoue}, {Madejski}, {Takahara}, {Tanihata}, \&
  {Kawai}}]{Kataoka-2001}
{Kataoka}, J., {Takahashi}, T., {Wagner}, S.~J., {et~al.} 2001, \apj, 560, 659

\bibitem[{{Kirk} \& {Mastichiadis}(1999)}]{Kirk-1999}
{Kirk}, J.~G. \& {Mastichiadis}, A. 1999, Astroparticle Physics, 11, 45

\bibitem[{{Larsson}(2012)}]{Larsson-2012}
{Larsson}, S. 2012, in Fermi \& Jansky 2011: Our Evolving Understanding of AGN

\bibitem[{{Mankuzhiyil} {et~al.}(2011){Mankuzhiyil}, {Ansoldi}, {Persic}, \&
  {Tavecchio}}]{2011ApJ...733...14M}
{Mankuzhiyil}, N., {Ansoldi}, S., {Persic}, M., \& {Tavecchio}, F. 2011, \apj,
  733, 14

\bibitem[{{Mastichiadis} {et~al.}(2013){Mastichiadis}, {Petropoulou}, \&
  {Dimitrakoudis}}]{Mastichiadis-2013}
{Mastichiadis}, A., {Petropoulou}, M., \& {Dimitrakoudis}, S. 2013, \mnras,
  434, 2684

\bibitem[{{Merrifield} \& {McHardy}(1994)}]{Merrifield-1994}
{Merrifield}, M.~R. \& {McHardy}, I.~M. 1994, \mnras, 271, 899

\bibitem[{{Nilsson} {et~al.}(2007){Nilsson}, {Pasanen}, {Takalo}, {Lindfors},
  {Berdyugin}, {Ciprini}, \& {Pforr}}]{Nilsson-2007}
{Nilsson}, K., {Pasanen}, M., {Takalo}, L.~O., {et~al.} 2007, \aap, 475, 199

\bibitem[{{Papadakis} \& {Lawrence}(1993)}]{Papadakis-1993}
{Papadakis}, I.~E. \& {Lawrence}, A. 1993, \mnras, 261, 612

\bibitem[{{Pichel, A. for the VERITAS Collaboration}(2009)}]{Pichel-2009}
{Pichel, A. for the VERITAS Collaboration}. 2009, arXiv:astro-ph/0908.0010

\bibitem[{{Poutanen} {et~al.}(2008){Poutanen}, {Zdziarski}, \&
  {Ibragimov}}]{Poutanen-2008}
{Poutanen}, J., {Zdziarski}, A.~A., \& {Ibragimov}, A. 2008, \mnras, 389, 1427

\bibitem[{{Punch} {et~al.}(1992){Punch}, {Akerlof}, {Cawley}, {Chantell},
  {Fegan}, {Fennell}, {Gaidos}, {Hagan}, {Hillas}, {Jiang}, {Kerrick}, {Lamb},
  {Lawrence}, {Lewis}, {Meyer}, {Mohanty}, {O'Flaherty}, {Reynolds}, {Rovero},
  {Schubnell}, {Sembroski}, {Weekes}, \& {Wilson}}]{Punch-1992}
{Punch}, M., {Akerlof}, C.~W., {Cawley}, M.~F., {et~al.} 1992, \nat, 358, 477

\bibitem[{{Rieger} {et~al.}(2000){Rieger}, {Kirk}, \&
  {Mastichiadis}}]{Rieger-2000}
{Rieger}, F.~M., {Kirk}, J.~G., \& {Mastichiadis}, A. 2000,
  arXiv:astro-ph/0005479

\bibitem[{{Smith} \& {Vaughan}(2007)}]{Smith-2007}
{Smith}, R. \& {Vaughan}, S. 2007, \mnras, 375, 1479

\bibitem[{{Takahashi} {et~al.}(1996){Takahashi}, {Tashiro}, {Madejski}, {Kubo},
  {Kamae}, {Kataoka}, {Kii}, {Makino}, {Makishima}, \&
  {Yamasaki}}]{Takahashi-1996}
{Takahashi}, T., {Tashiro}, M., {Madejski}, G., {et~al.} 1996, \apjl, 470, L89

\bibitem[{{Timmer} \& {Koenig}(1995)}]{Timmer-1995}
{Timmer}, J. \& {Koenig}, M. 1995, \aap, 300, 707

\bibitem[{{Tramacere} {et~al.}(2009){Tramacere}, {Giommi}, {Perri},
  {Verrecchia}, \& {Tosti}}]{Tramacere-2009}
{Tramacere}, A., {Giommi}, P., {Perri}, M., {Verrecchia}, F., \& {Tosti}, G.
  2009, \aap, 501, 879

\bibitem[{{Uttley} {et~al.}(2003){Uttley}, {Edelson}, {McHardy}, {Peterson}, \&
  {Markowitz}}]{Uttley-2003}
{Uttley}, P., {Edelson}, R., {McHardy}, I.~M., {Peterson}, B.~M., \&
  {Markowitz}, A. 2003, \apjl, 584, L53

\bibitem[{{Uttley} {et~al.}(2002){Uttley}, {McHardy}, \&
  {Papadakis}}]{Uttley-2002}
{Uttley}, P., {McHardy}, I.~M., \& {Papadakis}, I.~E. 2002, \mnras, 332, 231

\bibitem[{{Vaughan} {et~al.}(2003){Vaughan}, {Edelson}, {Warwick}, \&
  {Uttley}}]{Vaughan-2003}
{Vaughan}, S., {Edelson}, R., {Warwick}, R.~S., \& {Uttley}, P. 2003, \mnras,
  345, 1271

\end{thebibliography}
\begin{appendix}

\section{Reliability of $F_{\mathrm{var}}$}

\begin{figure}
  \includegraphics[width=88mm]{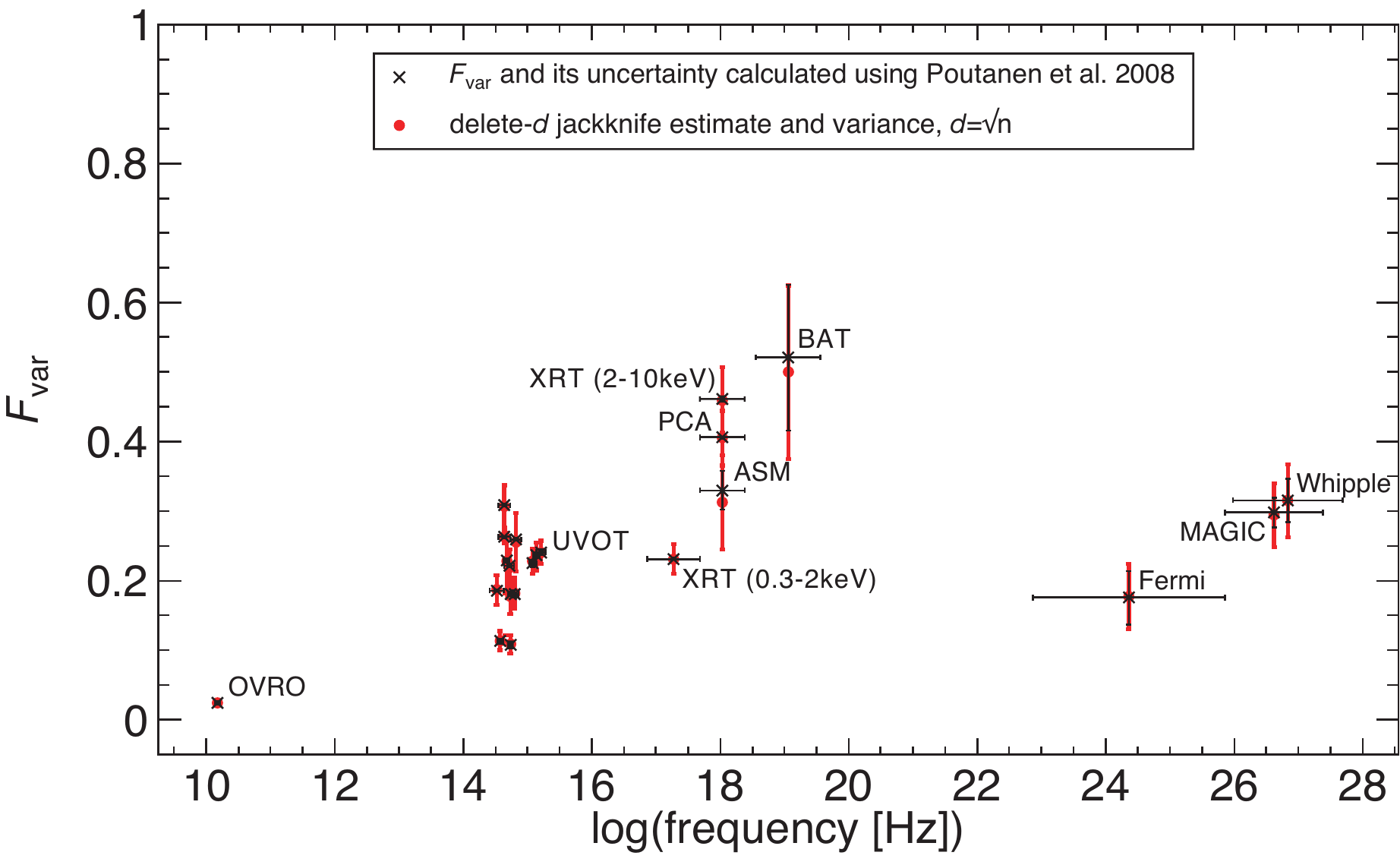}
  \caption{Measured fractional variability $F_{var}$ (black) compared with a delete-$d$ jackknife
   estimate (red) with $d=\sqrt{N}$, for all light curves with a minimum of $15$ data points.} %
\label{fig:jackknife}
\end{figure}

Almost all light curves of the campaign are unevenly sampled. The
sampling is different for each instrument depending on observation
schedule, weather and technical issues. There are often gaps of
different lengths in the light curves and each light curve has a
different number of data points ranging from a few up to a few
hundred. In addition, some light curves are binned into bins of several days
because of the limited sensitivity of the corresponding
instruments. Therefore we have to assess whether there is an error
introduced to $F_{\mathrm{var}}$ by the uneven sampling and the
binning and how the $F_{\mathrm{var}}$ values can be compared. In addition we need
to know the minimum number of flux measurements per light curve which
are needed to obtain a reliable, unbiased $F_{\mathrm{var}}$ value.

\begin{figure*}
  \includegraphics[width=17cm]{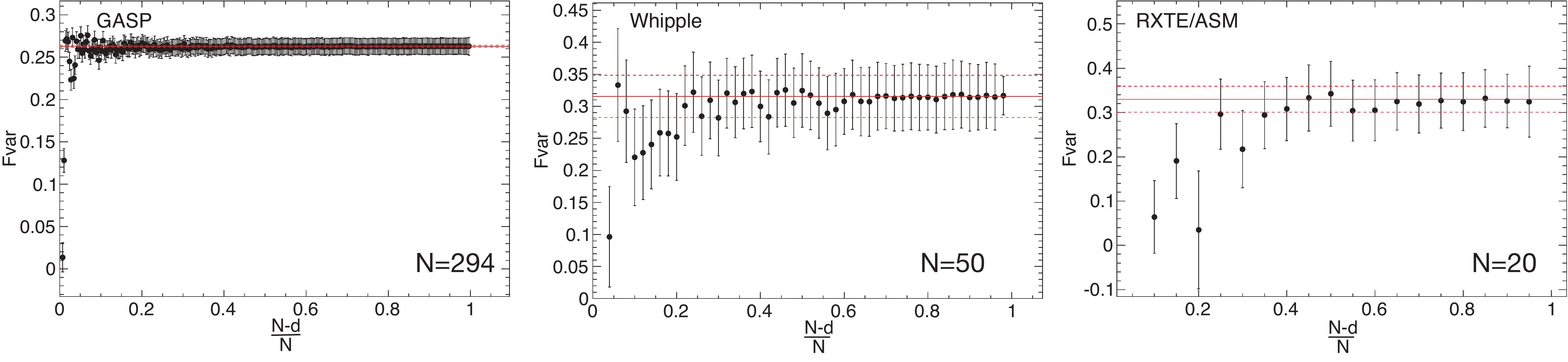}
  \caption{Fractional variability $F_{var}$ (red) as a function of $(N-d)/N$
    for the jackknife-$d$ samples of selected representative
    light curves. The measured $F_{var}$ of the original light curve and
    its error are shown as red horizontal lines. The $F_{var}$ of the
    jackknife-samples is constant and agrees with the original
    $F_{var}$ within its errors for all but the largest $d$ (i.e., the
    smallest jackknife-samples).
}
\label{fig:fvar_vs_d}
\end{figure*}

To address these questions, we first made a delete-$d$ jackknife
analysis, i.e., from each light curve containing $N$ data points, we
randomly removed $d$ data points ($1\leq d\leq N-2$) and calculated
$F_{\mathrm{var}}$ for this reduced sample of $N-d$ data points.
Applying a delete-$d$ jackknife analysis on a time series is formally
not correct, as the data are correlated in time (blazar light curves
usually show a red-noise behavior). However, our purpose is to create
datasets that have statistical properties identical to the real data
to demonstrate the impact of gaps and uneven sampling on
$F_{\mathrm{var}}$.  Figure \ref{fig:jackknife} shows
$F_{\mathrm{var}}$ for the jackknife datasets with $d=\sqrt{N}$
in comparison with $F_{\mathrm{var}}$ for the original
light curves. The $F_{\mathrm{var}}$ values do not change
significantly. The error bars are larger because of the reduced number
of flux values in the jackknife samples. The result does not depend on
the particular choice of $d$, as long as there are sufficient
datapoints in the jackknife-samples remaining. Some of the light curves
are (almost) regularly sampled, namely \emph{Fermi}-LAT, RXTE/PCA and
RXTE/ASM. These are good examples that demonstrate that irregular
sampling does not introduce a bias to the $F_{\mathrm{var}}$
measurement.

We also varied $d$ between $1$ and $N-2$. Figure \ref{fig:fvar_vs_d}
shows $F_{\mathrm{var}}$ vs. $(N-d)/N$ for selected light curves. As long
as $N-d$ is larger than $\sim5$, the measured $F_{\mathrm{var}}$ is
approximately constant with varying $d$ and in agreement with
$F_{\mathrm{var}}$ of the original light curve. Likewise, the error bars do
not change significantly. Strong deviations from $F_{\mathrm{var}}$ of
the original light curve occur only, if at all, when $N-d<10$. No
significant deviations are observed at $N-d\geq20$ for any of the
light curves. Thus we conclude that our $F_{\mathrm{var}}$ measurement
is robust for all light curves with $20$ or more flux data points. In
our multi-wavelength sample most light curves in the optical, UV,
X-rays, HE $\gamma$ rays and VHE have more than $20$ flux data points. In
the radio and near-infrared, most light curves do not.

$F_{\mathrm{var}}$ is reliable for all but the smallest samples. If after removal of $d$ data points 
the remaining sample is smaller than about $10$, then $F_{\mathrm{var}}$ might be over- or 
underestimated.

\begin{figure}
  \includegraphics[width=88mm]{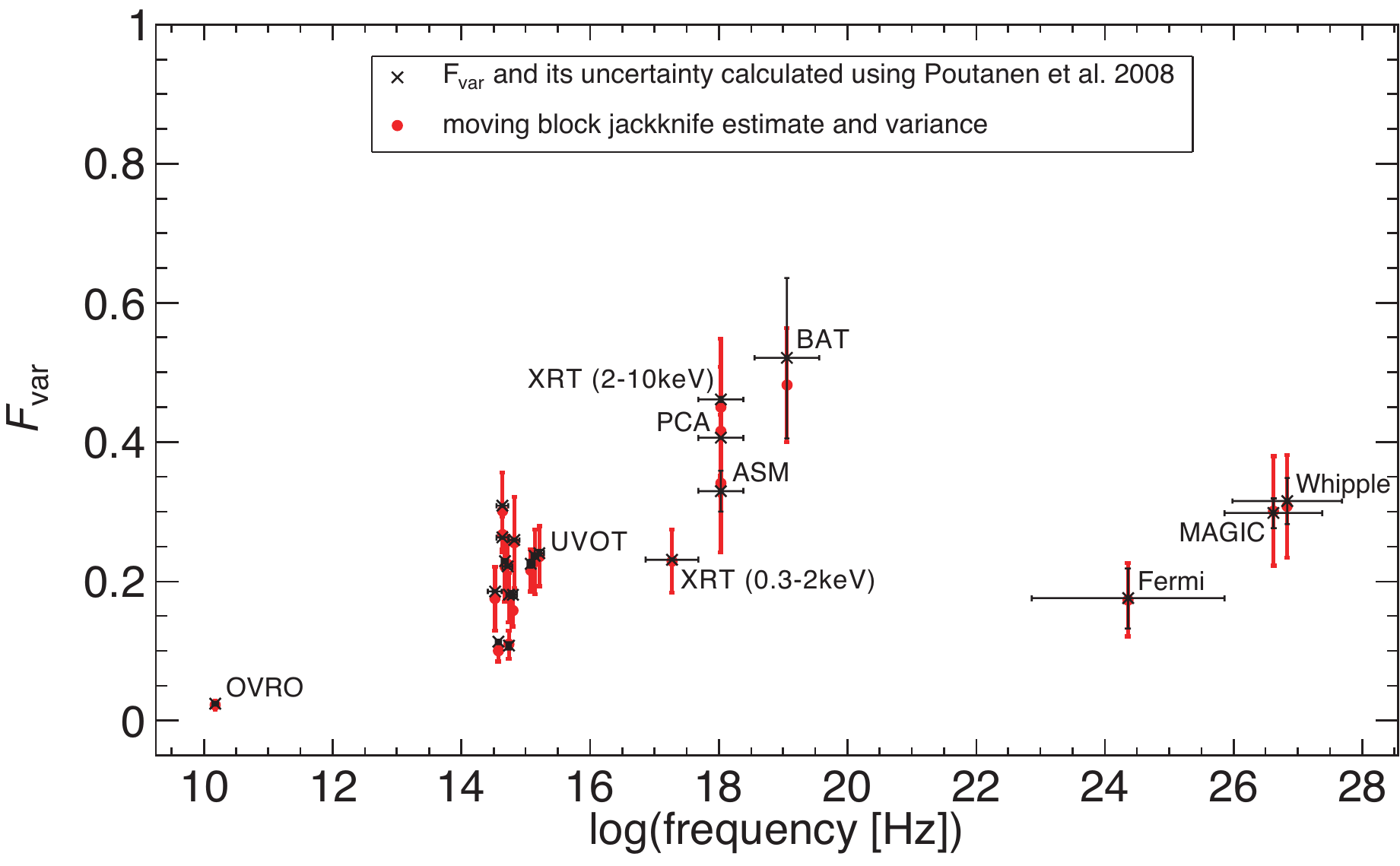}
  \caption{Measured fractional variability $F_{var}$ (black) compared with a moving-block jackknife
   estimate (red), using a blocksize of $m=\sqrt[3]{N}$, for all light curves with a minimum of $15$ 
   data points.}
\label{fig:jackknife-block}
\end{figure}

We also made a moving-block jackknife test, i.e., we removed blocks of
$m$ consecutive measurements. This test is still formally correct when
the data are slightly correlated in time, but the drawback is that the
number of jackknife samples is much smaller than in case of the
delete-$d$ jackknife test. Likewise, it only shows the influence of gaps,
not the influence of a random sampling. Figure \ref{fig:jackknife-block}
shows the test for $m=\sqrt[3]{N}$. As in the case of the delete-$d$
jackknife test, the $F_{\mathrm{var}}$ values do not change
significantly and the uncertainties are larger.

These conclusions, however, are only valid because the dataset does
not have any strong flare and hence it is unaffected by removing
points or blocks randomly. Occasional strong flares therefore should
be removed before doing an $F_{\mathrm{var}}$ analysis of light curves
that are otherwise in a typical or low state.

\end{appendix}

\end{document}